# Three-Dimensional Nanotransmission Lines at Optical Frequencies: A Recipe for Broadband Negative-Refraction Optical Metamaterials


*Andrea Alù and Nader Engheta*

*University of Pennsylvania*

*Department of Electrical and Systems Engineering*

*Philadelphia, PA 19104, U.S.A.*



## Abstract

Here we apply the optical nanocircuit concepts to design and analyze in detail a three-dimensional (3-D) plasmonic nanotransmission line network that may act as a negative-refraction broadband metamaterial at infrared and optical frequencies. After discussing the heuristic concepts at the basis of our theory, we show full-wave analytical results of the expected behavior of such materials, which show increased bandwidth and relative robustness to losses. The possibility and constraints of getting a 3-D fully isotropic response is also explored and conditions for minimal losses and increased bandwidth are discussed. Full-wave analytical results for some design examples employing realistic plasmonic materials at infrared and optical frequencies are also presented, and a case of sub-wavelength imaging system using a slab of this material is numerically investigated.




## *1. Introduction*

The interest in artificial materials and metamaterials is currently growing in the scientific communities, as many research groups all over the world are working on this topic and numerous research articles have been devoted to this subject. The possibility of mimicking nature in its way of tailoring the material response to electromagnetic and optical waves with appropriate choices of the "molecules" that compose a bulk material has attracted scientists since the end of the 19[th] century. At microwave frequencies, metamaterials made of artificial conducting or dielectric small inclusions with specific shapes has been applied to lenses, filters and antennas. In this way, novel material responses not readily available at the frequency of interest may be tailored for various applications. In particular, in this range of relatively low frequencies the possibility of building a double-negative (DNG) material, i.e., a medium with real parts of permittivity and permeability simultaneously negative at the same frequency, has recently been demonstrated experimentally [1], by embedding in a host dielectric small resonant inclusions with electric and magnetic response. As first envisioned in [2], a planar slab of such a material, in the ideal limit of no losses, may have the outstanding property of focusing the image of an object with its sub-wavelength details. This is one of the many anomalous properties that such a material, characterized by backward plane wave propagation and an effective negative index of refraction [3], may show over a limited frequency band, which may be useful for applications in many areas. The intrinsic resonances of each electric and magnetic inclusion, required in this configuration to get such anomalous effects, however



imply several limitations, both in terms of bandwidth and of robustness to ohmic losses and to other imperfections. Moreover, building isotropic inclusions with a behavior independent of the polarization and direction of the impinging wave has been a challenging task in practice.

The extension of these ideas to higher frequencies, possibly up to the visible domain, meets additional challenges, partly due to the fact that the metallic inclusions generally used at lower frequencies lose their conducting properties in the infrared and visible frequencies [4]. A mere scaling of the inclusion shape has been shown to have certain limitations beyond 100-200 THz [5]-[6]. Other ways of synthesizing such resonant inclusions at higher frequencies have been proposed in recent times, usually exploiting the resonance of the circulating displacement current rather than of the conduction current, but similar limitations on bandwidth and losses are evident also in this case (see e.g., [7]-[8] and references therein).

Another alternative way of building metamaterials at microwaves is to employ lumped circuit elements, connected in a "right-handed (RH)" or "left-handed (LH)" transmission line (TL) configuration in order to obtain the desired forward- or backward-wave propagation property. This setup has been demonstrated by several groups to be effective in the 2-D planar case for creating a microwave TL metamaterial with effective LH properties. Furthermore, the sub-wavelength focusing and other anomalous wave interaction properties have been investigated experimentally for this configuration [9]-[12]. An extension to the 3-D TL configuration has been expectedly more challenging in terms of construction, but the same concept has been recently applied successfully to synthesize in principle



a 3-D circuit network with isotropic left-handed properties [13]-[15]. An experimental realization of such 3D network has also been recently presented [16]. These solutions do not rely on the resonances of the single elements composing the material, but rather on the pass-band properties of a cascade of inductors and capacitors properly arranged, which, as in any TL setup, offers a wider frequency band of operation, due to the tight coupling of the single resonances between each TL cell. Therefore, the bandwidth of operation and the relative robustness to the possible presence of losses and other imperfections in this case are higher compared with the techniques that employ resonant inclusions for synthesizing metamaterials. The above-mentioned 3-D microwave circuit setup, however, experiences similar constraints when the design is to be scaled up in frequency: due to the variation of the material conduction at higher frequencies (such as IR and visible domains), no lumped circuit elements have been available in the usual conventional sense in the THz, infrared and higher frequency regimes, and therefore such TL networks have not been previously attempted at such high frequencies.

In a recent paper, however, we have envisioned a different paradigm for synthesizing lumped nanocircuits at infrared and optical frequencies, making use of arrangements of plasmonic and non-plasmonic particles that may respectively act as lumped nanoinductors and nanocapacitors [17]. In that work we have shown how, for an external observer, a single dielectric nanoparticle, small compared to the operating wavelength, may effectively be regarded as a nanocapacitor, relating the averaged voltage across its volume with the



displacement current ($\partial D / \partial t$, where $D = \varepsilon E$ is the displacement vector and $E$ is the local electric field) through an equivalent capacitively reactive impedance proportional to its permittivity $\varepsilon$. For a plasmonic particle, whose real part of permittivity is negative at the frequency of interest [18], the corresponding negative capacitance may be treated as an equivalent nanoinductance, and this allows envisioning more complex nanocircuits by "connecting" such basic nanoelements. Following this idea, in [19] we have speculated that cascades of plasmonic and non-plasmonic nanoparticles may act as 1-D or 2-D nano-TL at infrared and optical frequencies, and we have shown how actually this analogy applies to layered plasmonic and non-plasmonic materials. In particular, in the limit in which such cascades of nanocircuit elements are properly arranged and the gap among them goes to zero, we have theoretically shown how such a nano-TL may be envisioned as thin planar layers (2-D) or cylindrical rods (1-D) of plasmonic and non-plasmonic materials, and how their operation at optical frequencies may be regarded, by all means, analogously to the conventional cascades of series lumped inductors and shunt lumped capacitors (right-handed TL) or series capacitors and shunt inductors (left-handed TL) [19]. In that work we have investigated analytically how their operation is analogous in many ways. In other words, such thin layers or rods of plasmonic and non-plasmonic materials act at optical frequencies the same way as TL would do at lower frequencies.

Here we analyze the possibility of applying these concepts to synthesize a three-dimensional (3D) negative-index metamaterial at optical frequencies by properly "connecting" plasmonic and non-plasmonic particles in a 3-D network, acting



respectively as nanoinductors and nanocapacitors. In this way we may heuristically envision a negative-index 3-D nano-TL (isotropic or anisotropic) metamaterial at optical frequencies with the advantages of relatively broad bandwidth and relatively low sensitivity to losses, which is typical of this class of TL microwave metamaterials, but now in the optical regime. The availability of natural plasmonic materials with reasonably low losses at THz, infrared and optical frequencies is well known [18], and the classes of noble metals (silver, gold, aluminum…), polar dielectrics (silicon carbide, lithium tantalate, zinc telluride…), and some semiconductors (indium antimonide [20]…) are examples of such natural materials. Their anomalous interaction with light has been known for decades and widely reported in the literature.

Constructing a negative-index metamaterial at optical frequencies is an important step in expanding the exciting features of metamaterials into the optical domains, providing potential applications in several fields, spanning optics, microscopy, biology, and communications, to name a few. To date, a few experimental groups have demonstrated the presence of negative refraction and/or negative magnetic response at optical frequencies [21]-[27]. One of the main goals in the experimental efforts remains to be the lower sensitivity to losses and the higher bandwidth for these effects. Although frequency dispersion and presence of losses are necessary in passive materials with such anomalous properties [4], we predict here that the introduction of the TL concepts at these high frequencies may represent a viable solution to the problem of realizing a 3-D negative-index optical metamaterial with relatively lower loss and higher bandwidth, consistent



with the analogous achievements at microwave frequencies. Following a similar train of thoughts for applying the nanocircuit concepts presented in [17] and [19] to their microwave TL metamaterials, Grbic and Eleftheriades have recently presented some preliminary results on their efforts to synthesize a 3-D isotropic optical negative-index material [28], as first anticipated heuristically in another paper from their group [29]. In their geometry, however, the sizes of inclusions are not much smaller than the wavelength and, as a result, the effective negative refraction is influenced by the lattice resonances in the periodic metamaterial -- similar in some senses to the results reported in [8]. Furthermore, a relatively "large" size of the basic inclusion constrains the fundamental lower limit to the resolution for sub-wavelength imaging applications [30]. As we show in the following, in our approach the inclusions are assumed to be much smaller than the wavelength, which results in tight coupling of these "nanocircuit elements" and the possibility of supporting a nano-TL mode with broader bandwidth for the negative refraction.

A small portion of the results reported in the present manuscript has been presented orally in a recent symposium [31]. An $e^{-i\omega t}$ time dependence is assumed throughout the manuscript.

## 2.      Heuristic Nano-Circuit Analogy

A right-handed 1-D TL is made of a cascade of series inductors and shunt capacitors, as schematically represented in Fig. 1 (top left). In its discrete representation this is a low-pass filter that supports forward-wave propagation



along its axis of propagation below its cut-off frequency. Exchanging the role of inductors and capacitors one synthesizes a left-handed 1-D TL (Fig. 1, top right), which is a high-pass filter and it supports backward-wave propagation above its cut-off frequency, with phase and energy flowing in anti-parallel directions, as in a DNG material [3]. The unit cells of 3-D version of these TLs are depicted in Fig. 1, middle row. Again, the right-handed one (Fig. 1, middle left), originally introduced by Kron in 1943 [32], would support forward-wave modes in the three directions, whereas the left-handed dual configuration (Fig. 1, middle right), suggested by Grbic and Eleftheriades [13], would support isotropic backward-wave propagation. When the unit cell dimensions tend to zero, the operational bandwidth of such 3-D networks in principle tends to infinity, due to the coupled resonances among the cascaded infinitesimal cells, similar to what happens in the 1-D version, for which the cut-off frequencies tend respectively to infinity and zero. As originally recognized in [32], propagation in a dielectric with positive constitutive parameters may indeed be interpreted as the voltage and current propagation along the network of Fig. 1 (middle, left) with infinitesimally small cell size, where the loops of series inductors couple with the impinging orthogonal magnetic field and provide the effective permeability, whereas the shunt capacitors couple with the parallel electric field and provide the effective permittivity. Such a 3-D circuit network would therefore behave, in the limit of electrically small unit cells, as an effective "double-positive" material with isotropic forward-wave propagation in all directions. Reversing the role of inductors and capacitors, as suggested by Grbic and Eleftheriades [13], is



equivalent to flipping the signs of permittivity and permeability of the effective medium, since a negative inductor (capacitor) at a specified frequency acts as an effective capacitor (inductor). This is how the 3-D version of the planar TL negative-index metamaterial has been envisioned at microwave frequencies [13]-[16], providing larger bandwidth and better robustness to losses than the double-negative metamaterial made of resonant inclusions, as discussed in the introduction.

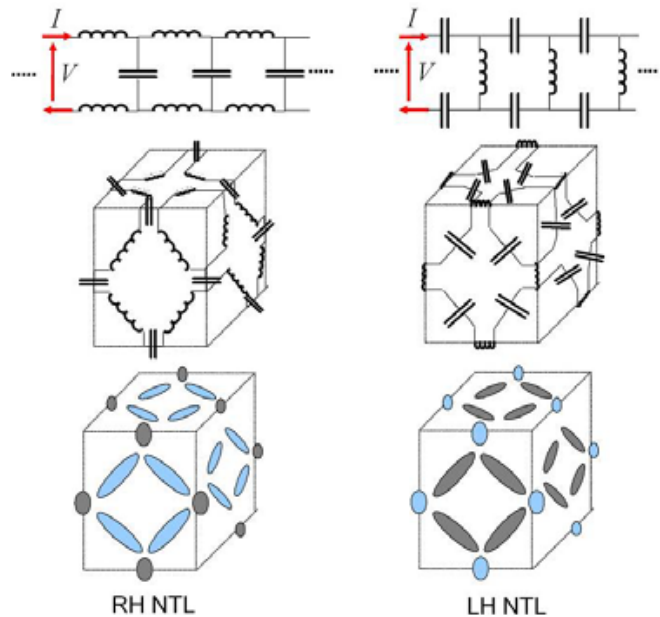

RH NTL   LH NTL

Figure 1 – (Color online). The concept of nano-TL (NTL) to build a 3-D isotropic negative-index material at optical frequencies. Top row: 1-D circuit model for a TL in its right-handed (left) and left-handed (right) configuration; Middle row: The unit cell of analogous 3-D version, to build an isotropic TL double-positive [32] (left) or double-negative (right) metamaterial at microwave frequencies, as suggested in [13]; Bottom row: The unit cell of optical version of the 3-D TL metamaterials by applying the concepts of nano-circuit elements presented in [17], i.e., utilizing plasmonic (blue, lighter) and non-plasmonic (grey, darker) nanoparticles as nanoinductors and nanocapacitors, respectively.



As done in the bottom row of Fig. 1, applying the circuit analogy presented in [17], one can substitute every inductor with a plasmonic (blue, lighter) particle and every capacitor with a non-plasmonic (grey, darker) dielectric particle, effectively synthesizing heuristically a right-handed or left-handed 3-D TL medium at optical frequencies. In the following, we analyze whether this heuristic analogy may hold, and therefore whether it is possible to extend the concepts of 3-D isotropic TL metamaterials to the optical regime. In our vision, this may provide a novel possibility for design of isotropic and anisotropic effective negative-index materials at optical frequencies with relatively large bandwidth and more robustness to losses.

### 3.    *Full-Wave Dispersion Relation*

A nanoparticle with a size much smaller than the wavelength of operation is well described in terms of its effective dipolar polarizability $\alpha$, which relates the impinging local electric field $\mathbf{E}$ to its induced dipole moment $\mathbf{p}$ as $\mathbf{p} = \alpha\mathbf{E}$. We assume for simplicity that the particle polarizability is isotropic, which may be obtained with spherical or 3-D symmetric particles, but this analysis may also be extended to anisotropic tensorial polarizabilities, or we may assume it valid for a specific orientation of the impinging field.

The 3-D left-handed nano-TL medium of Fig. 1 (bottom right) may be more easily envisioned as a medium composed of plasmonic nanoparticles, of polarizability $\alpha$, interleaved by gaps in a background medium with permittivity $\varepsilon_0 > 0$. Following the previous section discussion, we therefore claim, as will be



shown in our analysis below, that a regular lattice of densely packed and properly designed plasmonic particles may support a nano-TL propagating mode and thus may act as an effective negative-index metamaterial, even though no direct magnetic response is present in the materials under consideration (i.e., at the frequencies of interest all the magnetic permeabilities of the materials are equal to $\mu_0$, the one of free space, and due to the small size of the inclusions the magnetic polarizability of each of them is negligible). The dual lattice, i.e., a dense array of "voids" in a plasmonic host medium, would correspond to a double-positive metamaterial with forward-wave behavior.

The propagation properties in such a medium may be obtained analytically by considering the eigenvalue problem associated with this crystal lattice: supposing that a plane wave propagates along this collection of particles with phase vector $\boldsymbol{\beta} = \beta_x \hat{\mathbf{x}} + \beta_y \hat{\mathbf{y}} + \beta_z \hat{\mathbf{z}}$, the associated induced dipole moments in the crystal will be given by $\mathbf{p}_{lmn} = \mathbf{p}_{000} e^{i\boldsymbol{\beta}\cdot\mathbf{r}_{lmn}}$, with $\mathbf{r}_{lmn} = l\, d_x \hat{\mathbf{x}} + m\, d_y \hat{\mathbf{y}} + n\, d_z \hat{\mathbf{z}}$ being the position of the particle with index $(l, m, n)$. Here, without loss of generality, we have assumed the principal axes of the lattice to be oriented with the Cartesian coordinates and $l, m, n$ being any integer number.

The eigenmodes supported by this medium satisfy the equation:

$$\sum_{(l,m,m)\neq(0,0,0)} \underline{\mathbf{G}}\left(\mathbf{r}_{lmn}\right) \cdot \mathbf{p}_{lmn} = \alpha^{-1} \mathbf{p}_{000}, \qquad (1)$$

where $\underline{\mathbf{G}}$ is the dyadic Green's function of an electric dipole in the background medium [33]. This implies that such a mode can be self-sustained by the infinite regular array of dipoles, provided that $\boldsymbol{\beta}$ satisfies (1).



Due to the absence of magneto-electric coupling effects in the particles and in the background material, the eigenmodes supported by such a lattice are linearly polarized. This can be seen from the fact that if one assumes that in the left-hand side of Eq. (1) all the $\mathbf{p}_{lmn}$ are linearly polarized along one direction, then also the induced dipole moment $\mathbf{p}_{000}$ will be necessarily polarized in the same direction, due to the symmetry properties of $\underline{\mathbf{G}}$. This is independent of the direction of $\boldsymbol{\beta}$.

Due to this symmetry, the linearity of the problem and the arbitrariness of the periodicities along the three axes $d_x$, $d_y$, $d_z$, we may assume all the dipoles to be polarized along $\hat{\mathbf{x}}$, i.e., $\mathbf{p}_{000} = p\,\hat{\mathbf{x}}$. The eigenvalue equation (1) is slowly convergent, due to the propagating modes along the lattice, and it loses its convergence properties when complex $\boldsymbol{\beta}$ are considered. We have discussed and solved these convergence problems in the case of an infinitely long isolated chain of particles in [34]-[35] (which would correspond to the case in which Eq. (1) is limited to the summation over only one of the indices) by evaluating a closed-form solution and applying an analytical continuation, which may be similarly applied in this case. Moreover, several rapidly convergent solutions for lossless propagating modes (real-valued $\boldsymbol{\beta}$) in lossless lattices, which have been provided over the years for general 3-D scenarios [36]-[39], may be applied to the present analysis.

In the limit of no ohmic losses in the particles, for which the general relation $\mathrm{Im}\!\left[\alpha^{-1}\right] = -k_0^3 / 6\pi\varepsilon_0$ ($k_0$ being the background wave number) holds independent



of the nature of the particle [40], Eq. (1) becomes real valued [39] and it can be rewritten in the following normalized form:

$$\sum_{(l,m,m)\neq(0,0,0)} \mathrm{Re}\left(\frac{e^{i\left(\bar{r}+l\bar{\beta}_x\bar{d}_x+m\bar{\beta}_y\bar{d}_y+n\bar{\beta}_z\bar{d}_z\right)}}{\bar{r}^3}\left[\left(1-i\bar{r}\right)\frac{2l^2\bar{d}_x^2-m^2\bar{d}_y^2-n^2\bar{d}_z^2}{\bar{r}^2}+m^2\bar{d}_y^2+n^2\bar{d}_z^2\right]\right)=$$
$$=\frac{2}{3}\mathrm{Re}\left[\bar{\alpha}^{-1}\right]$$

$$(2)$$

where $\quad \bar{\boldsymbol{\beta}}=\boldsymbol{\beta}/k_0,\qquad \bar{d}=k_0d=2\pi d/\lambda_0,\qquad \bar{r}=\sqrt{l^2\bar{d}_x^2+m^2\bar{d}_y^2+n^2\bar{d}_z^2}\,,$

$\bar{\alpha}=k_0^3\alpha/\left(6\pi\varepsilon_0\right)$, and $\lambda_0$ is the background wavelength. This real dispersion relation, which is written in terms of normalized dimensionless quantities, relates the propagating modal solutions for a lossless lattice of particles embedded in a transparent background to the normalized geometrical and electromagnetic properties of the lattice. It is interesting to note how in this limit in which the dipolar response of the particle dominates, its properties are solely described by the quantity in the right-hand side of Eq. (2), which is just related to the geometrical and material properties of each inclusion composing the lattice. It should be underlined how Eq. (2) is an exact expression for the eigenmodes in such 3D array of particles, taking into account rigorously the dynamic interaction among all the particles, under the only assumption of considering each particle as a dipole polarizable entity. As we discuss more thoroughly in Section 6, this approximation yields accurate results in the limits we are interested here of electrically small plasmonic particles sufficiently close to their dipolar resonance.



Due to the periodicity of the system, the whole set of propagating modes is described by the real solutions with $\left| \bar{\beta}_i \right| < \pi / \bar{d}_i$, $i = x, y, z$. The fact that Eq. (1) is real valued, despite the presence of radiation loss mechanisms in each one of the lossless particles in the lattice described by the relation $\text{Im}\left[ \alpha^{-1} \right] = -k_0^3 / 6\pi\varepsilon_0$, is justified by the periodic configuration of the lattice, for which the radiation losses from each particle are simply "cancelled out" due to the destructive interference among the infinite number of them, as it may be verified analytically [39]. When small ohmic losses are considered, an analytic continuation of these solutions is required to give proper meaning to the summation in (1), analogously to what we have done for the 1-D configuration in [34]-[35]. We will discuss this point later in this paper. We underline here that under the dipolar approximation, the compact parameter $\text{Re}\left[ \alpha^{-1} \right]$, solely related to the geometrical and material properties of each inclusion, is sufficient to fully describe the electromagnetic eigenmodal properties of such complex metamaterial crystal, following (2). The $\text{Im}\left[ \alpha^{-1} \right]$, on the other hand, is related to energy issues, as already noticed in [35], and it will become relevant in Section 8, when ohmic losses in each particle are considered.

### 4.    *Nearest-Neighbor Approximation (NNA)*

Remaining in the lossless limit, it is instructive to analyze the nearest-neighbor approximation (NNA) of the eigenvalue problem (2), which, as we have shown in the 1-D case [34]-[35], yields approximately valid solutions in the limit of densely



packed lattices (with $\max\left(\overline{d}_x, \overline{d}_y, \overline{d}_z\right) \ll 1$) for sufficiently large values of $\overline{\beta}$. Considering in the summation only the two closest elements in each direction and taking the limit for small $\overline{d}$, we get the approximate expression for (2) as follows:

$$\frac{2\cos\left(\overline{\beta}_x \overline{d}_x\right)}{\overline{d}_x^3} - \frac{\cos\left(\overline{\beta}_y \overline{d}_y\right)}{\overline{d}_y^3} - \frac{\cos\left(\overline{\beta}_z \overline{d}_z\right)}{\overline{d}_z^3} = \frac{\operatorname{Re}\left[\overline{\alpha}^{-1}\right]}{3}. \qquad (3)$$

It is clear how under this approximation the propagation is decomposed into the three basic directions along the three axes of the lattice $(x, y, z)$, each of which analogous to the NNA in the case of 1-D linear configuration [34]-[35] (recalling that the polarization of the field has been fixed, i.e., $\mathbf{p}_{000} = p\,\hat{\mathbf{x}}$, Eq. (3) collapses to the NNA dispersion relations evaluated in [34]-[35] for $\overline{\boldsymbol{\beta}} = \overline{\beta}_x \hat{\mathbf{x}}$, $\overline{\boldsymbol{\beta}} = \overline{\beta}_y \hat{\mathbf{y}}$ or $\overline{\boldsymbol{\beta}} = \overline{\beta}_z \hat{\mathbf{z}}$). Indeed it turns out that independent of the nature of the particles, if a propagating mode is supported in this configuration and under this approximation, the TE propagation (with electric field orthogonal to the direction of propagation, i.e., with $\beta_x = 0$ in this case) corresponds to a backward-wave mode with negative group velocity for positive $\beta$, whereas a longitudinally polarized mode is necessarily forward wave. As shown analytically in [34]-[35], indeed the sign of the group velocity $v_g = \partial\omega/\partial\beta$ of the supported modes is always opposite to the sign of $\partial\operatorname{Re}\left[\overline{\alpha}^{-1}\right]/\partial\overline{\beta}$ in the corresponding dispersion relation. Since $\partial\operatorname{Re}\left[\overline{\alpha}^{-1}\right]/\partial\overline{\beta}_z > 0$, $\partial\operatorname{Re}\left[\overline{\alpha}^{-1}\right]/\partial\overline{\beta}_y > 0$ and $\partial\operatorname{Re}\left[\overline{\alpha}^{-1}\right]/\partial\overline{\beta}_x < 0$ in (3), the previous considerations hold. This confirms our heuristic model of Fig. 1: for TE



plane wave propagation, backward-wave modes are predicted by (3), consistent with the expected negative-index properties of such effective 3D metamaterial and with the left-handed TL analogy. In the longitudinal polarization, on the other hand, the roles of inductors and capacitors are effectively reversed, since the series (shunt) elements become effectively shunt (series) due to the different orientation of the electric field [35], and therefore the propagation behavior is expected to be reversed, i.e., the supported modes in this configuration are forward-wave modes. This is also consistent with the different behavior between the even and odd polarization of the fields in the planar nano-TL described in [19] and between the $n = 0$ and $n = 1$ modes guided along cylindrical nano-TL waveguides described in [41]. An extended paper thoroughly analyzing these analogies in different plasmonic geometries capable of supporting guided eigenmodes in light of their circuit model interpretation is under preparation in our group.

Provided that the lattice is sufficiently dense and that the wave number is sufficiently high (which corresponds to the requirements for supporting the nano-TL mode we are interested here, as shown in the following), the NNA agrees well with the exact solution for (2) and it confirms the TL analogy described in the previous paragraph. In particular, as we show in the following, we may get a relatively wide region of inverse polarizability values $\mathrm{Re}\left[ \overline{\alpha}^{-1} \right]$ (right-hand side of Eq. (2)-(3)) for which the metamaterial in many ways acts effectively as a DNG bulk material under transverse propagation, consistent with the broadband behavior of the TL approach. In this case, the plasmonic particles composing the



medium act as sub-wavelength "molecules", and their anomalous resonant behavior with light provides the necessary interaction for supporting a backward-wave, negative-index mode.

In the following we concentrate our analysis on the TE modes of propagation, which are those ensuring a negative-index behavior in this configuration, as the NNA predicts. It is interesting to underline, however, that for different applications it may be useful to consider the longitudinal modes of propagation, which may show other peculiar features, as we shortly discuss in the last section of the present manuscript. Moreover, when the roles of the background and inclusion materials are reversed, i.e., effectively synthesizing the dual nano-TL configurations, it is expected that such longitudinal modes are characterized by an effective negative index of refraction, and this may provide another interesting possibility for realizing double-negative metamaterials exploiting these nano-TL concepts. This will be the subject of future investigations.

### 5.    TE Propagation: Full-Wave Analytical Solution

The exact dispersion plots for $\overline{\boldsymbol{\beta}} = \overline{\beta}_y \hat{\mathbf{y}}$, i.e., for TE propagation, has been calculated by applying the acceleration technique from [39] for solving Eq. (2), and it is reported in Fig. 2 for different values of the lattice spacing. The figure shows several interesting features. First, in all cases for which the inverse polarizability $\mathrm{Re}\left[\overline{\alpha}^{-1}\right]$ of the particles is very high (that is when the particles weakly interact with the external field, they produce a weak scattering, and their $\alpha \to 0$) the values of supported $\overline{\beta}_y$ tend to unity, i.e., the supported modes are



TEM plane waves with phase velocity equal to that of the background material, consistent with what physically expected due to the absence of interaction with the particles. On the other hand, when the particle polarizability is near their individual resonance, i.e., for $\mathrm{Re}\left[\bar{\alpha}^{-1}\right] \simeq 0$, plane waves are strongly perturbed by the particles presence, and the possibility of backward-wave modes arises, consistent with the heuristic model of Fig. 1 and the NNA prediction in (3). Based on the previous discussion and consistent with the analysis in [34]-[35], backward-wave modes are supported in the regions where the slope of the curves of Fig. 2 is positive, i.e., where $\partial \bar{\beta} / \partial \mathrm{Re}\left[\bar{\alpha}^{-1}\right] > 0$, which happens only when $\bar{\beta}_y > 1$. The analytical proof of the dependence of the group velocity upon $\partial \bar{\beta} / \partial \mathrm{Re}\left[\bar{\alpha}^{-1}\right]$ is provided in the following section when the necessary presence of ohmic losses in the materials is considered. Physically this effect is justified by the fact that when $\bar{\beta}_y > 1$ the mode is confined around each longitudinal (along the propagation direction $y$) linear array of particles that composes the lattice, with all the other Floquet modes interacting among such chains being evanescent (notice that in order to have $\bar{\beta}_y > 1$, since $\left|\bar{\beta}_y\right| < \pi / \bar{d}_y$, then $d_y < \lambda_0 / 2$) [42]. We have shown analytically with closed form solutions [34]-[35], how transversely-polarized propagation along such an isolated chain of particles may be backward when $\bar{\beta}_y > 1$ and when the spacing between particles is sufficiently tight. This confirms the prediction in the general case of parallel arrays of such chains, together with the heuristic TL analogy of Fig. 1. In these negative-index regions



the behavior of $\bar{\beta}$ is monotonic and the agreement with the NNA is strong, particularly when the particles are tightly packed. One the other hand, in the regions of weak guidance, i.e., for lower values of $\bar{\beta}_y$, the mode is always a forward wave, as expected from the nature of the background material and from the fact that the supported mode is weakly guided by the particles. In this regime the NNA fails to predict the correct behavior, as expected.

Fig. 2a shows the case of densely packed particles, with $\bar{d}_y = 0.01$. When the lattice is fully isotropic, i.e., for $\bar{d}_x = \bar{d}_y = \bar{d}_z$, the interval of inverse polarizabilities for which the mode is backward is relatively narrow (notice however that the axis is normalized with respect to $\bar{d}_y^{-3}$) and slightly shifted towards positive values with respect to the resonance of the single particles. In order to increase such "bandwidth" of polarizabilities for the backward-wave operation, the longitudinal linear arrays should be moved farther apart from each other in the direction of polarization of the electric field, i.e., the distance $\bar{d}_x$ should increase. Since the coupling between them happens only through the Floquet evanescent modes in the backward region, when the distance $\bar{d}_x = 3\bar{d}_y$ the situation is already very similar to the case of an isolated chain [34], in which a wider range of inverse polarizabilities supports a backward-wave mode centered around $\mathrm{Re}\left[\bar{\alpha}^{-1}\right] \simeq 0$. The close proximity of the particles in the other transverse direction $\bar{d}_z$ interestingly increases further this bandwidth and shifts the resonance towards more negative values of $\mathrm{Re}\left[\bar{\alpha}^{-1}\right]$, as is evident from



comparison of the cases with $\bar{d}_z = \bar{d}_y$ and $\bar{d}_z = 3\bar{d}_y$, the last one of which has a similar dispersion to the isolated chain for clear reasons. We note how the value of $\bar{\beta}_y$ may become relatively large for having such backward waves, implying that the effective index of refraction of the bulk metamaterial may yield large negative values. In the upper part of the plot, i.e., for large values of $\bar{\beta}_y$, the effective wavelength inside the material becomes comparable with the distance between inclusions (the limit is $\bar{\beta}_y = \pi / \bar{d}_y$, for which the effective wavelength $\lambda_y = 2d_y$). In this limit, the concept of bulk material loses some of its meaning, and the effect may become somehow similar to a photonic band-gap structure, even though we still remain below the first lattice Bragg resonance. We point out that all the modes considered here still exist in the first Floquet band, for which the metamaterial may be indeed regarded as "homogeneous" and for which the negative index has a quasi-local meaning. In this negative-index regime, the TL coupling among the particles causes the anomalous behavior of the lattice. We discuss this in more detail in the following.

It should also be noted that for the curves with $\bar{d}_x = \bar{d}_y = \bar{d}_z$ the metamaterial is isotropic in 3-D, since the transverse propagation is allowed in all directions with analogous properties due to symmetry. This is an interesting outcome, since it may open up the possibility of manufacturing a relatively broadband and reasonably lower-loss negative-index isotropic metamaterial. In order to increase the bandwidth of operation, as we have mentioned above, it is possible to increase $\bar{d}_x$, i.e., the transverse direction along the polarization of the particles, and thus



breaking the isotropy of the medium in one direction. One of the interesting points from the analysis of Fig. 2 is that a closer coupling among the particles in the ($y-z$) plane transverse to the electric field polarization increases the bandwidth of operation and the robustness of the system, whereas coupling the particles in the direction of the electric field worsens these factors. This again may be heuristically explained in terms of the TL analogy, since the transverse or longitudinal geometrical disposition of the particles with respect to the field orientation may be interpreted as if they are cascaded respectively in "series" or in "parallel" with respect to the other neighboring particles. The tighter series cascade of such TL cells is expected to increase the overall bandwidth of operation, as it happens in a regular TL, whereas the higher coupling between different parallel lines closely packed together generally worsens the bandwidth and the robustness of the overall network.

Putting the particles farther apart in the propagation direction, as in Fig. 2b, confirms the same trend. Consistent with the isolated chain analysis [34], the boundaries of negative-index operation for the inverse polarizability do not change drastically when they are normalized to $\overline{d}_y^{-3}$.

When the distance between particles becomes even larger (Fig. 2c, $\overline{d}_y = 0.5$, i.e., $d_y = \lambda_0 / 4\pi$), the possibility of backward-wave modes is lost in the isotropic lattice configuration, even though increasing the distance among longitudinal chains with respect to the field orientation, i.e., increasing $\overline{d}_x$, backward-wave propagation is still possible. In the limit of an isolated chain, i.e., when $\overline{d}_x \to \infty$



and $\bar{d}_z \to \infty$, the limit of existence of the backward-wave mode is represented by

$\bar{d}_y = 1.517$ [35].

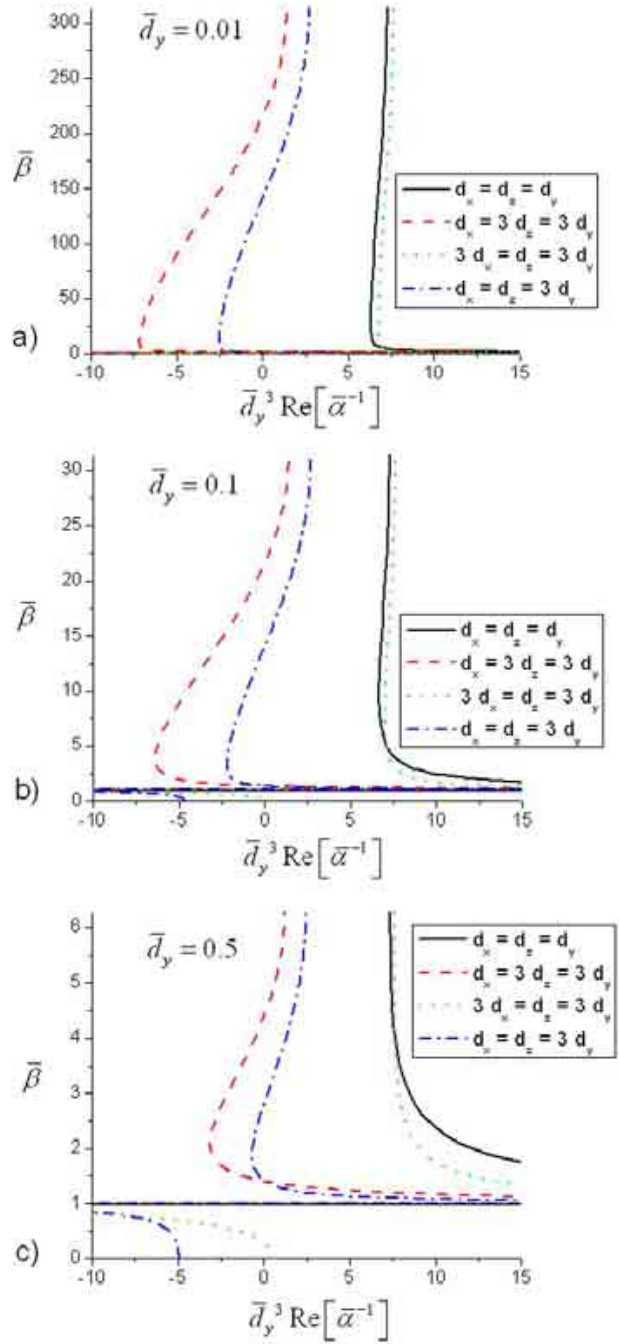



Figure 2 – (Color online). Normalized wave numbers versus normalized inverse polarizabilty for TE propagation (in this case $\beta_x = \beta_z = 0$ and $\mathbf{p}_{lmn} = p\, e^{im\beta_y d_y} \hat{\mathbf{x}}$ ) for different values of the distance between neighboring particles.

It is worth noting how the design of a 3D plasmonic negative-index metamaterials in [28] relies on a totally different phenomenon than the nano-TL mode considered here. To excite such a mode in an isotropic lattice the distance between neighboring particles needs to be much smaller than the wavelength in the background material, as Fig. 2 and the following results show. In [28], analogous to some other efforts exploiting plasmonic resonances of relatively larger particles (see e.g., [7]-[8], [43]), the negative-index effect is related to the simultaneous excitation of an electric and a magnetic plasmonic resonance, in the same band of frequencies, in each inclusion or in the lattice, whose period is often comparable with the wavelength in the background material. This effect relies on a physical phenomenon different from the one presented here. Our approach as presented here, which is based on the nano-TL theory introduced in the previous paragraphs, allows obtaining a relatively large bandwidth of operation despite the extremely small size of the individual cells composing the metamaterial. More discussion will be given in the following sections.

Figure 3 summarizes how the negative-index properties vary as a function of the lattice distances. As seen from Fig. 2, and consistent with the single chain propagation [34]-[35], the region of backward propagation is confined between a minimum value of $\overline{\beta}_y = \overline{\beta}_{\min}$ and the value $\overline{\beta}_y = \pi / \overline{d}_y$. Following the previous discussion on the relation between group velocity and the derivative



$\partial \mathrm{Re} \left[ \overline{\alpha}^{-1} \right] / \partial \overline{\beta}_y$, indeed it turns out that for these two extremes at which $v_g$ becomes null and flips its sign, the derivative $\partial \mathrm{Re} \left[ \overline{\alpha}^{-1} \right] / \partial \overline{\beta}_y = 0$.

The equation that is satisfied by $\overline{\beta}_{\min}$ is as follows:

$$\sum_{n=-\infty}^{\infty} \sum_{l=-\infty}^{\infty} \frac{\left( \overline{d}_x^{\,2} - 4l^2 \pi^2 \right) \sinh \left( \overline{d}_y f \right)}{f \left( \cos \left( \overline{d}_y \overline{\beta}_y \right) - \cosh \left( \overline{d}_y f \right) \right)^2} = 0, \qquad (4)$$

where $f = \sqrt{4\pi^2 \left( \dfrac{l^2}{\overline{d}_x^{\,2}} + \dfrac{n^2}{\overline{d}_z^{\,2}} \right) - 1}$. The series in (4) is rapidly convergent for its numerical evaluation.

In the interval $\overline{\beta}_y \in \left[ \overline{\beta}_{\min}, \pi / \overline{d}_y \right]$ the propagation is backward, and the plots for $\overline{\beta}_y = \overline{\beta}_{\min} \left( \overline{d}_y \right)$ are given in Fig. 3a. These lines, for different relative distances among the particles, represent the minimum wave number for getting a backward propagation (notice that all the curves are normalized to $\overline{d}_y^{\,-1}$), the upper value being $\overline{d}_y \overline{\beta}_y = \pi$. We note how in the isotropic case $\overline{d}_y \overline{\beta}_{\min} = \pi$ beyond $\overline{d}_y = 0.443$. This distance, which corresponds to $d_y = 0.07 \lambda_0$, represents the upper limit spacing to have backward propagation in an isotropic lattice supporting this nano-TL mode. This implies that a totally isotropic 3D nano-TL negative-index medium may be designed only with inclusions with total size of less than this value. Increasing the transverse spacing $\overline{d}_x$, the relative growth of $\overline{\beta}_{\min}$ is reduced, and even for a larger lattice spacing the backward-wave propagation is possible.



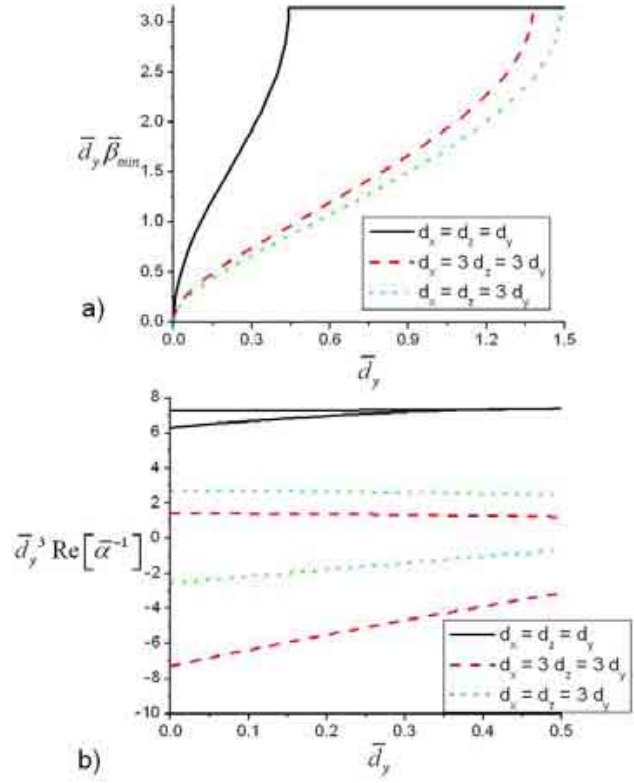

Figure 3 – (Color online). (a) Variation of the minimum value of $\bar{\beta}_y$ for achieving a backward-wave propagation, in terms of $\bar{d}_y$, for different values of the other spacing factors; (b) corresponding region of polarizabilities for which TE backward-wave propagation is supported.

Fig. 3b shows the corresponding admissible range of inverse polarizabilities (normalized to $\bar{d}_y^{-3}$) for having backward-wave propagation (note that the horizontal scale is smaller in this second plot, in order to zoom in this region for the isotropic case). The solid black lines delimit this region in the isotropic configuration, and it is seen how at $\bar{d}_y > 0.443$ the backward-wave propagation is lost. For larger $\bar{d}_x$, this region expands and shifts down, consistent with the results of Fig. 2. In this plot, the upper limiting line in every case corresponds to



the locus $\bar{\beta}_y = \pi / \bar{d}_y$. Decreasing the inverse polarizability, consistent with the backward behavior $\partial \operatorname{Re}\left[\bar{\alpha}^{-1}\right] / \partial \bar{\beta}_y > 0$, $\bar{\beta}_y$ decreases down to the bottom limit $\bar{\beta}_y = \bar{\beta}_{\min}\left(\bar{d}_y\right)$. At these two limits $\partial \operatorname{Re}\left[\bar{\alpha}^{-1}\right] / \partial \bar{\beta}_y = 0$, the group velocity is zero, consistent with the isolated chain configuration [34]-[35] (as we show in the following these two points are characterized by high attenuation factors when material losses are considered, therefore the meaning of group velocity may lose its practical meaning at these two points).

### 6. Plasmonic Particles

We now consider the case of plasmonic homogenous spherical particles of permittivity $\varepsilon$ for which an approximate expression for their inverse polarizability, adequate in the limit of small particles, is given by [18], [33]:

$$\operatorname{Re}\left[\bar{\alpha}^{-1}\right] = \frac{3}{2}\left(k_0 a\right)^{-3} \frac{\varepsilon + 2\varepsilon_0}{\varepsilon - \varepsilon_0}. \qquad (5)$$

Due to the geometrical requirement $d_y > 2a$ and the fact that $\dfrac{\varepsilon + 2\varepsilon_0}{\varepsilon - \varepsilon_0} > 1$ for any $\varepsilon > \varepsilon_0$, the lattice analyzed above cannot support any backward TL modes when it is made of standard dielectric particles with permittivity larger than that of the background medium, since $\bar{d}_y{}^3 \operatorname{Re}\left[\bar{\alpha}^{-1}\right] > 12$ for any $\varepsilon > \varepsilon_0$ (see Fig. 3). This is consistent with what we have found in the single chain configuration [34]-[35], it confirms the heuristic TL analogy that requires the presence of proper nanoinductors made of plasmonic materials to support propagation, and finally it is expected from the fact that such nano-particles need a strong interaction with



the local electric field in order to generate such anomalous behavior in the wave propagation, as guaranteed by their plasmonic resonant nature [18]. The region of permittivities for which negative-index operation is expected lies near the resonance of the particles, which is at around $\varepsilon = -2\varepsilon_0$ for the spherical particle. We can clearly shift this value by choosing differently shaped particles, for which the expression in (5) is modified.

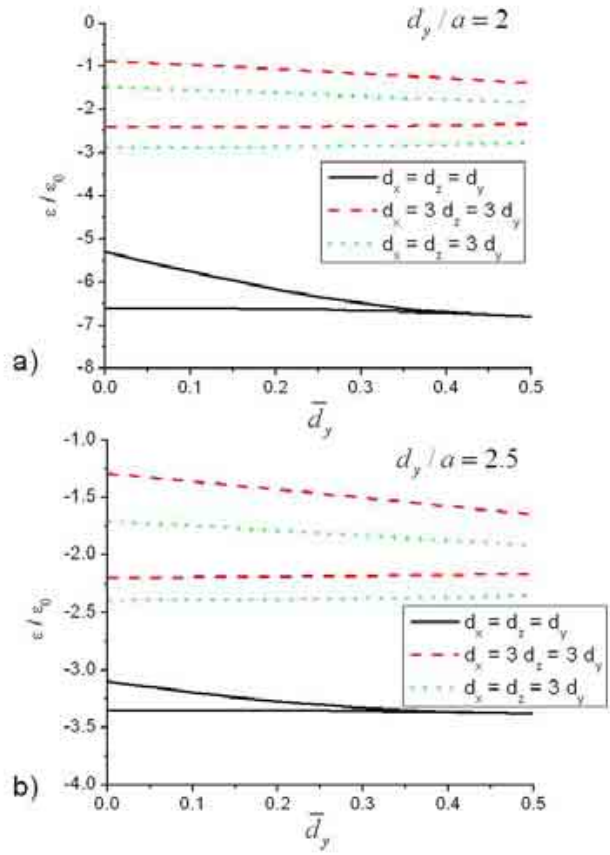

Figure 4 – (Color online). Regions of negative-index operation in terms of the permittivity of the spherical particle and their relative spacing factors in the three directions.

If in the negative-index region the inverse polarizability is bounded in the interval $f_1 < \bar{d}_y^{\,3} \operatorname{Re}\left[\bar{\alpha}^{-1}\right] < f_2$, obtained from Fig. 3b, then we can write this region in



terms of the permittivity $\varepsilon$ using (5), as we have similarly done for the 1-D linear chain [34]-[35]:

$$\frac{f_1/\eta+3}{f_1/\eta-3/2} < \varepsilon < \frac{f_2/\eta+3}{f_2/\eta-3/2}. \qquad (6)$$

Fig. 4 shows how the general results of Fig. 3 apply to the case of spherical homogeneous inclusions. It is apparent how by keeping fixed the geometrical ratio between the distance $d_y$ and the radius of the particle $a$, the bandwidth of operation increases for highly dense materials. This represents an interesting result, consistent with the 1-D configuration [34]-[35]: since the $Q$ resonance factor for an isolated particle increases in the quasi-static limit as $(k_0 a)^{-3}$, as in any dipolar radiating system [44], and its bandwidth decreases correspondingly, one would expect that by reducing the size of each of the inclusions composing the medium and therefore reducing the bandwidth of their individual resonance (which tightens up around the frequency for which $\varepsilon = -2\varepsilon_0$), the negative-index response of the whole metamaterial would also be reduced in bandwidth. This may be the case when metamaterials are designed as collections of resonant, but weakly coupled, inclusions. However, the possibility of packing a larger number density of particles and the strong coupling of the surrounding closely spaced inclusions in this nano-TL network allows the bulk material to resonate even at frequencies relatively far from the individual resonance of each inclusion. The result is a relatively broad range of permittivities for which backward-wave propagation is possible, and this value saturates to a finite range of permittivities even when the singe particle size tends to zero (of course at some point,



approaching the atomic size, quantum effects should be considered in the description of the particle interaction and this classic approach would not be adequate any more!). This counterintuitive result is again consistent with the TL analogy, for which even though each single cell composing the line resonates at a specific single frequency (Fig. 1 top), the overall line has a much larger (in principle infinite for infinitesimal cell size) bandwidth due to the strong mutual coupling among infinitesimally small, closely-packed cells in the cascade. This result is also confirmed by the fact that the range of admissible inverse polarizabilities diverges with $\bar{d}_y \to 0$ as $\bar{d}_y^{-3}$ (see Fig. 3b), consistent with the fact that when we are scaling down everything in dimensions, the inverse polarizability of the particle increases as $a^{-3}$ (5). These two geometrical factors compensate, and the region of admissible permittivities saturates to a finite, but fairly broad, range of values.

Due to the transformation rule (6), now the upper limiting line in every negative-index region of Fig. 4 refers to the limit $\bar{\beta}_y = \bar{\beta}_{\min}\left(\bar{d}_y\right)$, whereas at the lower limit of the backward-wave region, we have $\bar{\beta}_y = \pi / \bar{d}_y$, consistent with the facts that for backward-wave modes, $\partial \beta_y / \partial \omega < 0$ and for lossless particles $\partial \varepsilon / \partial \omega > 0$. Fig. 4a refers to the limit of touching spheres, i.e., $d / a = 2$, which is for geometrical reason the ideal maximum bandwidth achievable in terms of values of permittivities of the particles. It should be noted how in the isotropic lattice the resonance happens pretty far from the individual inclusion resonance ($\varepsilon = -2\varepsilon_0$) with a relatively broader bandwidth for smaller $\bar{d}_y$, but rapidly decreasing as



$\bar{d}_y \to 0.443$. Increasing the transverse spacing $\bar{d}_x$ brings the resonance to higher values of permittivities and broadens the bandwidth of operation. With a larger ratio $d/a = 2.5$, as in Fig. 4b, the decrease in the coupling between particles uniformly affects the performance in terms of bandwidth.

At this point it is suitable to digress for a moment in order to discuss the dipolar approximation in the description of the particle interaction that we are using in this analysis. As anticipated in Section 3, even though we take into account the full-wave dynamic coupling between the particles in an exact manner, we are still assuming in (1) that the field radiated by each particle is limited to the dipolar spherical harmonic, neglecting the higher-order multipolar contributions to radiation from each particle. Due to their small dimensions this assumption is valid at least in the far-field, as well established in [33]. In the very near-zone region of any particle, however, in general higher-order contributions of the particle itself may become dominant, since their field distribution grows as $r^{-2n-1}$, with $n$ being the multipolar scattering order. Since the particle is very small, near the surface of the particle its own field may contain other multipolar terms contributing to the interaction with neighboring particles. We claim, however, that this is not a significant issue for the present case, as discussed also in [35], since, as shown above, the negative-index propagation relies on the resonant behavior of the dipolar fields. This implies that amplitude of the excited resonant dipole moments are, in first approximation, independent of the size of the sphere, as shown in [45], whereas the other multipole contributions decrease with the size of the particle as $a^{2n+1}$, with $a$ being the averaged radius of the particle [45],



ensuring that for any $r > a$, independent of how small $a$ is and how close to the surface we are, the resonant dipolar scattering amplitude is dominant over any other non-resonant multipole contribution in this resonant configuration. We have verified the validity of this assumption with some numerical full-wave simulations (not shown here for the sake of brevity).

## 7. *Frequency Dispersion*

It is well known that assuming negative permittivity for passive materials implies dispersive response in frequency [4]. From Eq. (5), we get the following:

$$\frac{\partial \operatorname{Re}\left[\alpha^{-1}\right]}{\partial \omega} = -\frac{3a^{-3}}{4\pi} \frac{\partial \varepsilon}{\partial \omega} \left(\varepsilon_0 - \varepsilon\right)^{-2}, \quad (7)$$

which confirms that in regions of small ohmic absorption (clearly the only of interest here), where $\partial \varepsilon / \partial \omega > 0$ for any passive material [4], the inequality $\partial \operatorname{Re}\left[\alpha^{-1}\right] / \partial \omega < 0$ should hold. A glance at the monotonic behavior of $\beta$ in the regions where nano-TL propagation is supported (Fig. 3b and 4), together with these considerations, allows identifying such regions as backward-wave.

As an example, in Fig. 5 we consider a Drude-model dispersive material with permittivity $\varepsilon\left(\omega\right) = \varepsilon_0 \left(1 - 3 f_0^2 / f^2\right)$, i.e., for which the plasmonic resonance of the single nanosphere happens at frequency $f = f_0$. In the figure we have plotted the dispersion curves as a function of $f$ for a metamaterial made of such spheres with radius $a = \lambda_0 / 100$ at the frequency $f_0$ and interspacing distance (center to



center) $d_y = 2.1\,a$ , in order to verify the backward behavior, i.e., the negativity of $\partial\omega/\partial\beta$ .

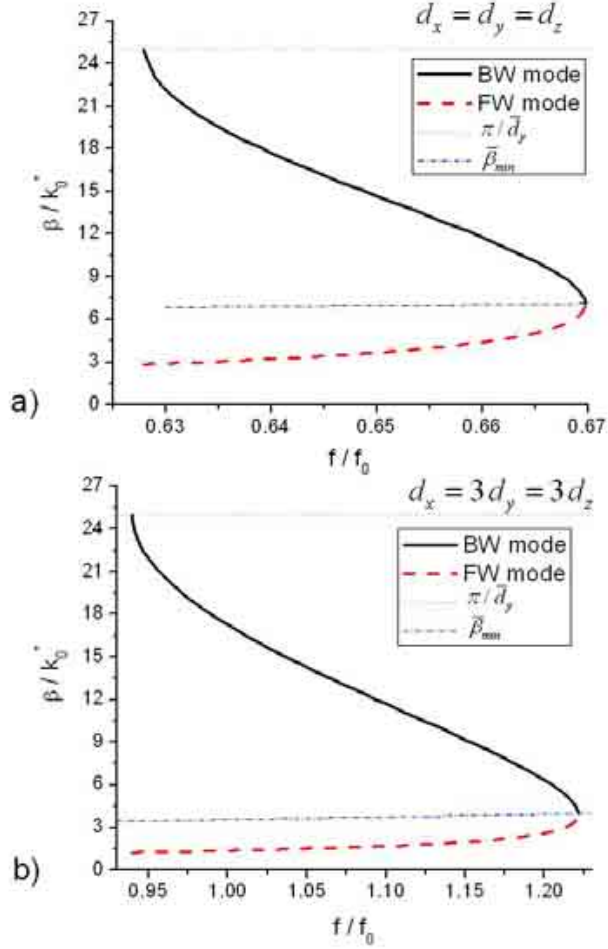

Figure 5 – (Color online). Normalized wave number $\beta$ (with respect to $k_0^* = 2\pi f_0\sqrt{\varepsilon_0\mu_0}$ ), vs. normalized frequency $f/f_0$, for TE propagation in a metamaterial made of plasmonic spheres with radius $a = \lambda_0/100$ (at frequency $f_0$), and center-to-center distance $d_y = 2.1\,a$ . The sphere permittivity is assumed to follow a Drude dispersion model with $\varepsilon(\omega) = \varepsilon_0\left(1 - 3f_0^2/f^2\right)$ .

The guided wave number along the lattice $\beta_y$ has been normalized in the plots to $k_0^*$ , i.e., the wave number in the background medium calculated at the frequency



$f_0$. The solid black line in the two cases of isotropic lattice and larger spacing along $x$ (respectively Fig. 5a and 5b) confirms the presence of a backward mode of propagation, following the left-handed nano-TL analogy, and allows envisioning a negative-index metamaterial over a relatively broad range of frequency. The line is bounded by the two limits predicted in Fig. 4 (the green dotted line for $\pi / \bar{d}_y$ and the blue dashed line for $\bar{\beta}_{\min}$, at whose intersections with the solid line the group velocity $\partial \omega / \partial \beta$ is zero. The dashed red lines correspond to the forward-wave mode of operation, as predicted in Fig. 2, which coexist together with the backward-wave mode over the whole negative-index regime, as we discuss in the following.

It is evident from the figures how the bandwidth of operation in the negative-index operation may be made relatively wide following this approach, and much wider than the one for the metamaterials employing resonant inclusions of such small dimensions, but weakly coupled. It is indeed interesting to note that while the resonance of the single nano-particle employed here has a very narrow bandwidth (its $Q$ factor for the case at hand is $Q = 1.5 \cdot 10^6$, while its fractional bandwidth is approximately proportional to the inverse of $Q$), the strong coupling among these resonant particles markedly widens the bandwidth of the metamaterial in its bulk, as evident from the figures. In Fig. 5a, for the isotropic case, a fractional bandwidth of backward operation of about 6.5% has been obtained from our analysis, which is increased to 26% when $\bar{d}_x$ is increased. In the isotropic lattice, as predicted by the previous figures, the resonance is also shifted down with respect to the resonant frequency of the individual particles,

$f_0$, and by increasing $\overline{d}_x$ the central frequency of negative-index operation is shifted back to the position in which each particle resonates individually. Again, the transverse coupling among the different longitudinal lines in the direction of electric polarization affects dramatically the overall resonance.

The birefringence of this type of medium is quite evident from the plots of Fig. 5, as well as the previous dispersion figures, implying that at a given frequency two transverse modes with different phase velocities would propagate in the same direction. Interestingly enough, one of them is backward and the other one consists of a forward wave. The forward one is only weakly affected by the presence of the scatterers and is widespread over the background material, with its phase velocity being only slightly lower than that of the background itself. The backward-wave mode is instead tightly packed to the particles in the direction of propagation, consistent with the analysis of the 1-D scenario [35], and this may provide ways of isolating this mode from the forward one for some potential applications. We are currently working on these issues.

Although a complete physical analysis of the anomalous dispersion predicted in such a metamaterial is out of the scope of the present manuscript, here we note that the intrinsic birefringence of this medium is indeed a symptom of the strong spatial dispersion present in such a dense lattice of plasmonic particles near their dipolar resonance. The birefringence due to the possible anisotropy of a material would usually be associated with different polarizations of the field for the different supported eigenmodes; however, here two different modes are shown to coexist with the electric field polarized transversely to the direction of



polarization, in addition to a third mode with longitudinal polarization, as predicted by the NNA. Not to mention that this anomalous dispersion is present even when $d_x = d_y = d_z$ (as in Fig. 5a) – condition that ensures absence of anisotropy in the metamaterial. Due to the strong interaction among such resonant small plasmonic inclusions positioned in the very near-field of each other, upon which the negative-index nano-TL propagation is based, the assumption that the local averaged polarization of the material depends only on the local applied field and is not affected by the contributions from other more distant particles appears to be less applicable here. Such non-local phenomena here, which arise also naturally for instance in crystals [46], are associated with an intrinsic spatial dispersion of the metamaterial under analysis. This is particularly evident noticing that a spatially dispersive crystal indeed supports two transversely polarized and a longitudinally polarized wave [46], consistently with the results of the present analysis.

In general, therefore, a plane wave impinging on a slab made of such a metamaterial would excite more than one eigenmode propagating inside the slab, even for normal incidence, where the two transverse modes will be in principle both excited. Their relative amplitude of excitation will depend on the boundary conditions and on the form of excitation, noting that this boundary-value problem requires to employ additional boundary conditions, extracted from the physics of the problem at the interfaces of the slab [46]-[47]. This is however beyond the present discussion and it will be analyzed in the near future by our group in a more extensive study of this specific problem. It should be stressed here,



however, the necessity of suppressing the possible excitation of the other eigenmodes when the interest relies only on the negative-index behavior of such a metamaterial. As already mentioned, one way of doing this may rely in the peculiar field distribution of this nano-TL mode, which, due to the high wave-number, is mainly concentrated around the particles, distinct from the other modes supported by the lattice.

On a different note, it is interesting to underline the possibility of moving the resonant frequency of the medium by introducing more degrees of freedom in this problem. An example can be given by the use of concentric core-shell spherical particles made of two different materials. As we have shown in [45], the employment of combinations of plasmonic and non-plasmonic materials allows tailoring the frequency of polarizability resonance of such nanoparticles by choosing different filling ratios for such core-shell particles. This implies that it may be possible to employ a material whose permittivity at the desired frequency may not fall into the range of the admissible values required in Fig. 4 to get the proper negative-index behavior, but by filling part of the spherical inclusions with a non-plasmonic material, one may bring the resonant frequency into this desired region. Still at least one of the two materials should be plasmonic in order to excite a notable resonance in such a small volume and the consequent negative-index propagation in the composite metamaterial.

Consider for instance the example of Fig. 5 at the working frequency $f_0 = 600 \, THz$ for which $\lambda_0 = 500 \, nm$. Each particle has a diameter $2a = 10 \, nm$ and the lattice spacing is $d_y = d_z = d_x / 3 = 12 \, nm$, with the background medium



being free space. When the spheres are homogeneous (i.e., the particles consists of a single material) the variation of the normalized wave number $\beta$ in terms of the particle permittivity at the frequency $f_0$ is reported in Fig. 6a, where one can appreciate the range of permittivities in which the metamaterial would support a backward-wave mode. In this example, the range of permittivities for a backward-wave propagation falls in the range $-2.217 < \varepsilon < -1.332$. The band of frequencies for which a realistic material may be available with these required values of permittivities may not be centered at the frequency of interest, and it may not necessarily correspond to a low-loss region for the material itself. Imagine, instead, that we are willing to use for this same geometry particles made of silver, which at $\lambda_0 = 500\,nm$ have a real part of permittivity given by $\mathrm{Re}\left[\varepsilon_{Ag}\right] = -9.77\varepsilon_0$ [48]. Silver has a relatively low ohmic absorption at this frequency, so it looks to be a suitable material for these purposes. However, the real part of permittivity clearly does not fall in the range of allowable permittivities for desired backward-wave propagation. Covering however spheres of silicon carbide ($\varepsilon_{SiC} = 6.52\varepsilon_0$ [49], with radius $a_1 = \eta\,a$) with shells of silver, while maintaining the same outer radius $a$ for the core-shell particles, we can move the resonance of the particles to the desired frequency, as we have shown in [45] for different purposes. Following the same full-wave analysis, it is possible to calculate the dispersion of $\beta$ with the ratio of radii $\eta = a_1 / a$ at the frequency of interest, i.e., with $\lambda_0 = 500\,nm$. This is done in Fig. 6b, (assuming for the moment zero material loss for the silver and SiC), where it is noticeable how it is



indeed possible to select the desired backward-wave behavior at the frequency of interest by properly choosing the filling ratio of the spheres.

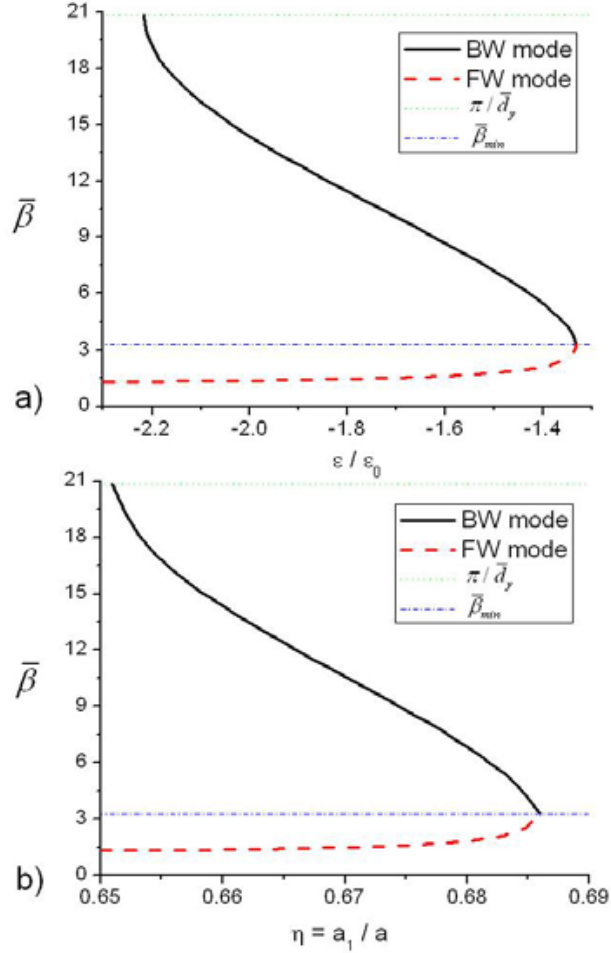

Figure 6 – (Color online). Variation of the normalized wave number $\bar{\beta}$: (a) with the permittivity of homogeneous spherical particle of radius $a = 5\,nm$; (b) with the ratio of radii for concentric core-shell spherical particle with inner core (radius $a_1$) made of $SiC$, with $\varepsilon_{SiC} = 6.52\varepsilon_0$ and outer shell (radius $a = 5\,nm$) made of silver with $\mathrm{Re}\left[\varepsilon_{Ag}\right] = -9.77\varepsilon_0$ (here not considering ohmic losses). In these plots $d_y = 2.1a$ and the operating wavelength (in the host medium) is $\lambda_0 = 500\,nm$.

## 8.    *Presence of Material Losses*



Even though we have shown up to this point how the TL concepts may be applied to realize an optical left-handed metamaterial, allowing to increase its bandwidth of negative-index operation, an important step towards the practical realization of devices based on these concepts consists in realizing how the ohmic losses present in realistic materials may affect the previous considerations, and under which limits backward-wave propagation may be obtained in this metamaterial when material losses in its constituent inclusions are taken into account.

Material losses affect the polarizability of each particle by adding an imaginary part to the inverse of its value [35]:

$$\text{Im}\left[\overline{\alpha}^{-1}\right] = -1 - \overline{\alpha}_{loss}^{-1}, \qquad (8)$$

where the limit of $\text{Im}\left[\overline{\alpha}^{-1}\right] = -1$ holds only for lossless particles and takes into account of the radiation mechanism. Introducing a non-zero $\overline{\alpha}_{loss}^{-1}$ (which is necessarily positive for passive media), Eq. (1) is no longer a real-valued equation, and the supported wave numbers have now complex solutions, which can be found after an analytical continuation of the series in (1). In the limit of small losses, such solutions consist of a perturbation of the lossless case, as we have shown for the 1-D linear chain in [35]. By expanding in Taylor series to the first order Eq. (1), in fact, it is possible to show that the complex solution for $\overline{\beta} = \overline{\beta}_r + i\,\overline{\beta}_i$ is given by:

$$\overline{\beta}_i = -\overline{\alpha}_{loss}^{-1} \frac{\partial \overline{\beta}_r}{\partial \text{Re}\left[\overline{\alpha}^{-1}\right]}, \qquad (9)$$



where $\bar{\beta}_r$ still satisfies Eq. (2) neglecting the loss. This formula is accurate as long as we consider reasonably low loss tangent factors in the materials, consistent with the realistic values of natural plasmonic materials, such as noble metals, at visible frequencies. Notice how Eq. (9), due to the sign of $\bar{\beta}_i$, supports the previous statement that forward (backward) modes are characterized by the condition $\partial \bar{\beta}_r / \partial \mathrm{Re}\left[\bar{\alpha}^{-1}\right] < 0$ ($\partial \bar{\beta}_r / \partial \mathrm{Re}\left[\bar{\alpha}^{-1}\right] > 0$). This is due to the fact that causality requires a mode to decay in the direction of power propagation when material losses are present. Since we assumed an $e^{i\boldsymbol{\beta} \cdot \mathbf{r}}$ propagation and positive $\bar{\beta}_r$ in deriving (9), and since $\bar{\alpha}_{loss}^{-1} > 0$ in passive materials, it follows that positive $\beta_i$ would correspond to forward-wave modes, with power and phase propagation in the same direction, whereas negative $\beta_i$ would correspond to backward-wave modes.

As seen from (9), the attenuation factor $\bar{\beta}_i$ is linearly proportional to the ohmic factor $\bar{\alpha}_{loss}^{-1}$ through the derivative $\partial \bar{\beta}_r / \partial \mathrm{Re}\left[\bar{\alpha}^{-1}\right]$. This implies that the regions near the cut-off of the backward-wave mode ($\bar{\beta} = \bar{\beta}_{\min}$ or $\bar{\beta} = \pi / \bar{d}$), where the group velocity is near zero, are the most affected by the presence of losses, whereas the regions farther away from these extremes are less affected by the presence of losses in the inclusions. As a corollary, a larger bandwidth implies the possibility of regions where the derivative in (9) is lower, and therefore losses may affect less the wave propagation.

The condition of minimum losses for given material properties, and therefore fixed $\overline{\alpha}_{loss}^{-1}$, may therefore be derived by requiring minimum $\partial \overline{\beta}_r / \partial \operatorname{Re}\left[\overline{\alpha}^{-1}\right]$, which gives:

$$\sum_{n=-\infty}^{\infty} \sum_{n=-\infty}^{\infty} \frac{\left(\overline{d}_x^{\,2} - 4l^2\pi^2\right)\left(-3 + \cos\left(2\overline{d}_y\overline{\beta}_y\right) + 2\cos\left(\overline{d}_y\overline{\beta}_y\right)\cosh\left(\overline{d}_y f\right)\right)\sinh\left(\overline{d}_y f\right)}{f\left(\cos\left(\overline{d}_y\overline{\beta}_y\right) - \cosh\left(\overline{d}_y f\right)\right)^3} = 0 ,$$

$$(10)$$

whose rapidly convergent solutions $\overline{\beta}_{md}$ (the subscript standing for "minimum damping") represent the optimum choices for guided modes in this negative-index material. Fig. 7 shows how this loss factor affects the propagation when varying the geometry of the lattice. Fig. 7a, in particular, shows the variation of the position of $\overline{\beta}_{md}$ changing the geometry of the lattice. We notice how at the cut-off of the modes $\overline{d}_y\overline{\beta}_{md} = \pi$ and for lower spacing the product $\overline{d}_y\overline{\beta}_{md}$ converges to a finite value. Fig. 7b shows the minimum attenuation factor $\partial \overline{\beta}_r / \partial \operatorname{Re}\left[\overline{\alpha}^{-1}\right] = \left|\overline{\beta}_i\right| / \overline{\alpha}_{loss}^{-1}$ obtained by varying the spacing $\overline{d}_y$ and using $\overline{\beta}_y = \overline{\beta}_{md}$, and therefore shows the variation of the minimum absorption factor with respect to the spacing in the lattice for fixed $\overline{\alpha}_{loss}^{-1}$. Again, for denser lattices, together with an increase in bandwidth, it is also possible to get less sensitivity to losses.

It should be underlined, however, that for geometrical reasons a smaller spacing between the centers of neighboring particles necessarily implies a smaller available volume for the single particle composing the lattice, and the value of



$\overline{\alpha}_{loss}^{-1}$ generally increases with a decrease in the volume of the particle for a fixed material loss.

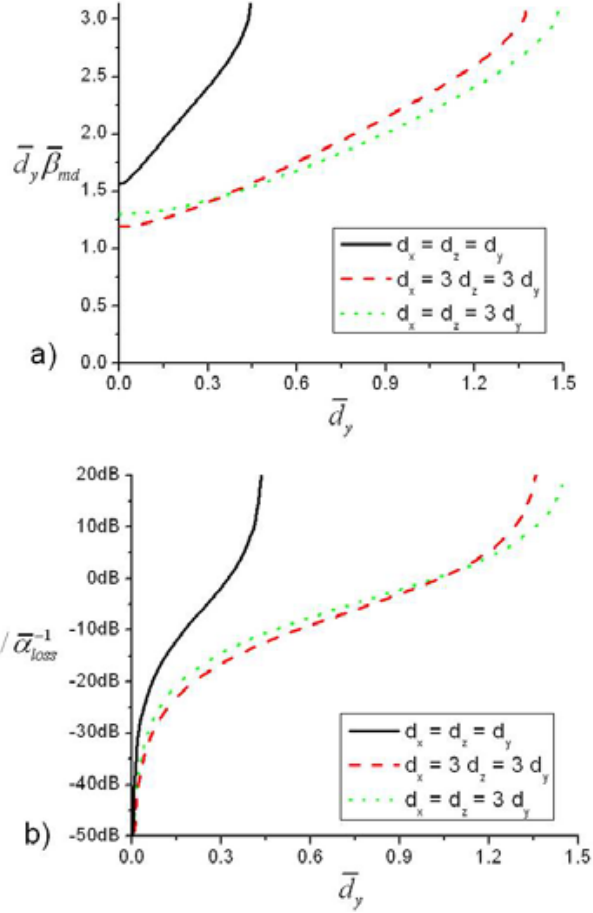

Figure 7 – (Color online). (a) Behavior of $\overline{\beta}_{md}$ (satisfying Eq. (10), normalized to $\overline{d}_y^{-1}$) versus the normalized longitudinal distance $\overline{d}_y$ between neighboring particles; (b) Normalized attenuation factor, $\overline{\beta}_i / \overline{\alpha}_{loss}^{-1}$, for $\overline{\beta}_y = \overline{\beta}_{md}$, i.e., variation of the minimum attenuation factor for a given material loss factor $\overline{\alpha}_{loss}^{-1}$.

This effect may be considered manipulating Eq. (5), since for a homogeneous particle with complex permittivity $\varepsilon = \varepsilon_r + i\varepsilon_i$, we can write [35]:



$$\overline{\alpha}_{loss}^{-1} = \frac{9\varepsilon_i}{2} \frac{\varepsilon_0 (k_0 a)^{-3}}{(\varepsilon - \varepsilon_0)^2 + \varepsilon_i^2}. \qquad (11)$$

If on the one hand $\overline{\alpha}_{loss}^{-1}$ grows with the imaginary part of the permittivity, as physically guessable, Eq. (11) shows that it also increases with the inverse volume of the particle. Therefore, if reducing the spacing among the particles may on the one hand increase the bandwidth performance of the negative-index propagation, together with the factor $\partial \overline{\beta}_r / \partial \mathrm{Re}\left[\overline{\alpha}^{-1}\right] = \left|\overline{\beta}_i\right| / \overline{\alpha}_{loss}^{-1}$, the corresponding reduction in the size of the particles increases the value of $\overline{\alpha}_{loss}^{-1}$, representing a lower limit for squeezing the lattice compactness when realistic losses are considered. Since the increase in $\overline{\alpha}_{loss}^{-1}$ in the sub-wavelength limit is proportional to $(k_0 a)^{-3}$, faster than the decrease in $\partial \overline{\beta}_r / \partial \mathrm{Re}\left[\overline{\alpha}^{-1}\right] = \left|\overline{\beta}_i\right| / \overline{\alpha}_{loss}^{-1}$, there is a trade-off between bandwidth of the metamaterial on one side and sensitivity to losses on the other when choosing the proper spacing (and corresponding dimensions of the inclusions) in the lattice. The important point to underline, as evident from the previous analysis, is that the particles embedded in the medium for a fixed size should be sufficiently dense in the transverse plane (i.e., transverse with respect to the polarization vector of the electric field) in order to get optimal performance in terms of bandwidth and sensitivity to losses.

Figure 8 shows the behavior of the curves in Fig. 7b when multiplied by $(k_0 a)^{-3}$, supposing $a = d_y / 2.1$. As you can see there is an optimum spacing where the sensitivity to losses is minimum, since for too closely packed lattices the ohmic absorption in the particles overcomes the benefits due to the coupling between



them, whereas a too-sparse lattice cannot guide backward-wave modes. In the isotropic lattice this minimum happens around $\bar{d}_y = 0.18$ ($d_y = 0.029\lambda_0$), as seen in Fig. 8a, whereas for larger spacing, as more evident in Fig. 8b that zooms into smaller values of the attenuation factor, the minima are placed at larger values of $\bar{d}_y$. This figure implies that "robustness" to losses may be obtained by accepting a trade-off in terms of bandwidth of negative-index operation, and that the isotropic setup is in general more sensitive to ohmic losses when compared with the anisotropic configuration with larger spacing in the direction of polarization of the field.

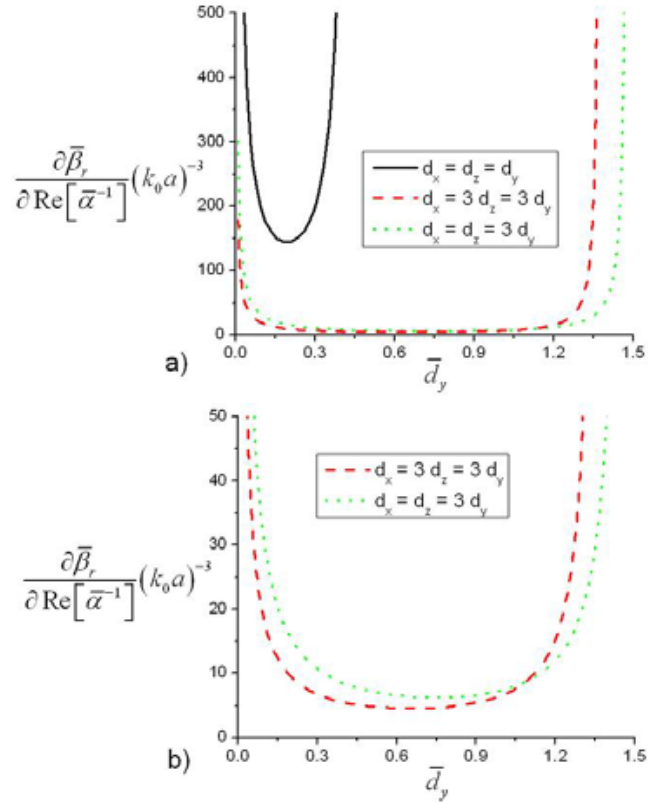

Figure 8 – (Color online). Attenuation factors (as in Fig. 7b, but normalized to the factor $(k_0 a)^3$ that takes into account of the increase in the loss coefficient when the particle size is reduced, as



determined in (11)) for $d_y = 2.1a$, showing how there exists an optimum spacing for obtaining the minimum sensitivity to losses.

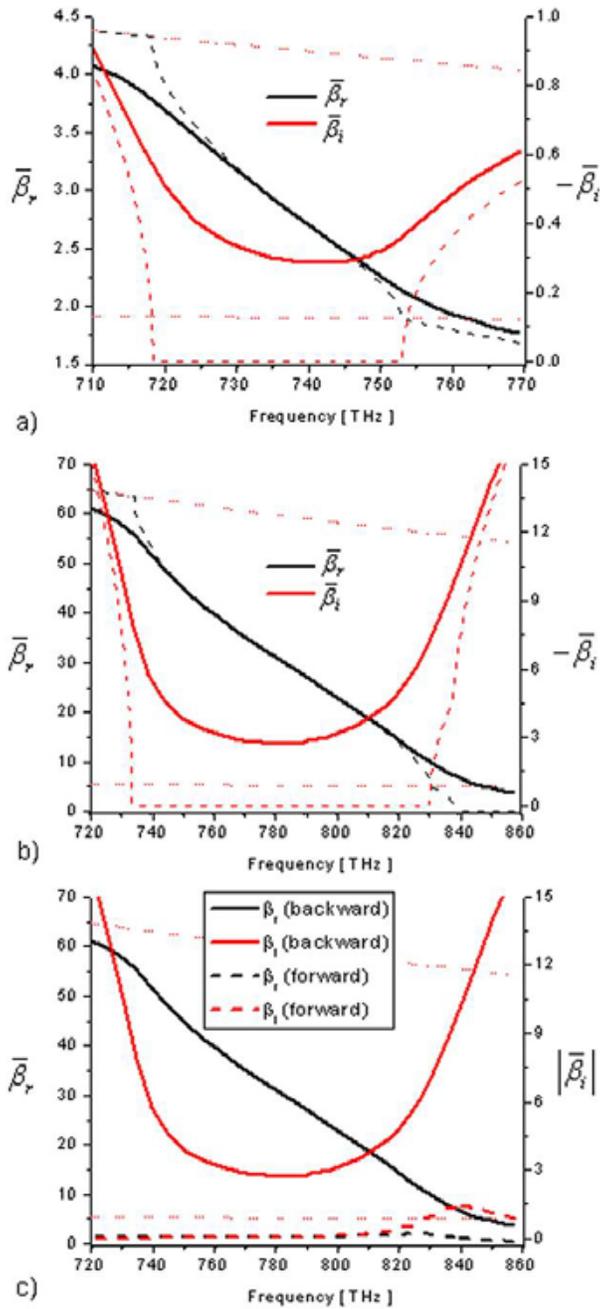

Figure 9 – (Color online). Dispersion plots of the negative-index operation versus frequency for a metamaterial made of silver particles, using silver material parameters (including losses) from experimental data for its bulk properties from [48], in a glass background ($\varepsilon_{SiO_2} = 2.14\varepsilon_0$). The



solid lines correspond to realistic data with losses, the dashed lines neglect the losses in the silver and the thin red dotted lines bound the region of backward-wave operation in the ideal lossless case. In (a): $a = 15\,nm$, $d_z = d_y = 33\,nm$, $d_x = 3d_y$; in (b) $a = 1\,nm$, $d_z = d_y = 2.2\,nm$; in (c) the backward-wave mode for Fig. 9b is compared to the forward-wave mode, showing the difference in the sensitivity to losses in the two cases. In (a) and (b), the forward-wave mode, although present, is not shown here to avoid crowding the plots.

It should be noted that the value of the attenuation factor $\bar{\beta}_i$ can be simply obtained by multiplying the value in the plot of Fig. 8 with the quantity $\dfrac{9\varepsilon_i\varepsilon_0}{2\left(\varepsilon_r - \varepsilon_0\right)^2 + \varepsilon_i^2}$. This quantity generally decreases with the frequency of operation when noble metals are considered (as it happens in the Drude or Lorentz model beyond its resonant frequency), therefore a resonance at lower frequencies, i.e., for more negative values of $\varepsilon_r$, is expected to be accompanied by a lower loss deterioration. This may be physically explained by the fact that the field hardly penetrates into the lossy plasmonic materials when their $\varepsilon_r$ is sufficiently negative, and this reduces the dissipated power.

### 9.    *Examples of Optical Negative-Index Metamaterials*

Consider as a first numerical example the design of the proposed left-handed metamaterial employing simple homogeneous silver particles. In this case we employ realistic data for the permittivity of bulk silver, including material dispersion and realistic material losses in the material, as taken from [48]. As a first case, Fig. 9a shows the frequency dispersion for a lattice of silver particles of



radius $a = 15\,nm$ and spacing between the particles $d_z = d_y = 33\,nm$ and $d_x = 3d_y$

in a glass background ($\varepsilon_{SiO_2} = 2.14\,\varepsilon_0$). In this case, the resonance $\mathrm{Re}\,\alpha^{-1} = 0$ is

for $f_0 \simeq 750\,THz$, $\lambda_0 \simeq 400\,nm$, and the spacing is $\bar{d}_y = 0.7$ at this frequency,

which is in the range of minimum losses (see Fig. 8). The solid lines report real

(black, darker) and imaginary (red, lighter) part of the wave number $\bar{\beta}_y$. The

dashed lines also show the behavior that such dispersion curves would have had,

if the silver losses in the particles had been neglected. The thin red dotted lines

plot the ideal boundaries that limit such dispersion curves, i.e., $\pi / \bar{d}_y$ and $\bar{\beta}_{min}$.

We see how the real part of the wave number is weakly affected by ohmic losses

except at the extremes of the guidance regions, where also the imaginary part

increases. This is consistent with the analytical predictions in the previous section.

For comparison, in Fig. 9b it is also reported the results for the same setup, but for

much smaller particles ($a = 1\,nm$) maintaining the same aspect ratio with the

spacing, i.e., $d_z = d_y = 2.2\,nm$. In this case $\bar{d}_y = 0.05$ at the central frequency. As

you can see comparing the two cases, the smaller spacing has definitely increased

the bandwidth of backward-wave behavior for this setup, together with a

corresponding increase in its loss factor, as predicted in the previous section. Also

the real part of $\bar{\beta}$ has increased, giving rise to a slower phase propagation, due to

the reduced distance between neighboring particles.

As already mentioned, in these plots the thin dotted lines delimit the region of

backward-wave operation in the lossless case, i.e., they are the loci where

$\bar{\beta}_r = \pi / \bar{d}_y$ and $\bar{\beta}_r = \bar{\beta}_{min}$. It is interesting to see how the points of zero group



velocity (infinite slope in the lossless curves when they meet the boundaries of backward-wave operation, as we have shown previously) are smoothed by the presence of losses and there is no longer a "point" at which the group velocity goes actually to zero in the lossy scenario, consistent with the linear chain configuration [35]. In the middle of the region of backward-wave operation, as underlined, the losses minimally affect the real part of the propagation constant $\bar{\beta}_r$, giving rise only to a negative imaginary part $\bar{\beta}_i$ (its negative value for positive $\bar{\beta}_r$ confirms the backward-wave nature of these modes). Entering the cut-off regions, i.e., when we are outside the region of negative-index operation, the attenuation factor is less affected by ohmic losses and $\bar{\beta}_i$ has a similar value to the lossless case, mainly caused by destructive interference among the particles (we have entered the stop-band region of the lattice). We note how in these first two plots (Fig. 9a and 9b) the forward-wave mode of operation has not been reported, in order to avoid crowding the plots. In Fig. 9c, for comparison, this mode has been added to the dispersion curves for lossy particles of Fig. 9b. As it may be appreciated, this forward-wave mode is characterized by a lower sensitivity to the material losses, as predicted by (9) and by the previous section analysis. This is clearly due to the fact that its interaction with the lossy particles is less strong. In the following we do not focus on this mode of operation, since it is not of interest for the present analysis.

Figure 10 shows analogous results obtained with a silicon carbide background ($\varepsilon_{SiC} = 6.52\varepsilon_0$ [49]). In this case, the frequency of operation is shifted down, since $\mathrm{Re}\,\alpha^{-1} = 0$ for $f_0 \simeq 550\,THz$, $\lambda_0 \simeq 550\,nm$. In Fig. 10a the radius of each particle



is fixed at $a = 10\,nm$ and spacing between the particles $d_z = d_y = 22\,nm$ and $d_x = 3d_y$. The normalized spacing in this case is again around $\bar{d}_y = 0.6$, ensuring minimum attenuation for this configuration. Employing smaller particles, as in Fig. 10b, i.e., $a = 1\,nm$, $d_y = 2.2\,nm$ and $\bar{d}_y \approx 0.06$, the bandwidth is sensibly increased, but also the attenuation factors are higher, consistent with the previous section and Fig. 9. The use of lower loss materials may help in this sense to find a good trade off between bandwidth and sensitivity to losses.

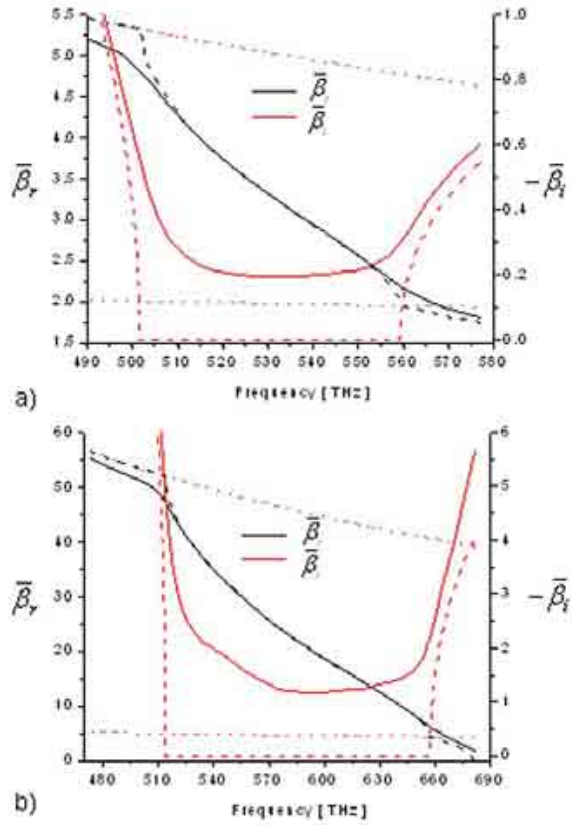

Figure 10 – (Color online). Similar to Fig. 9, but with a $SiC$ background ($\varepsilon_{SiC} = 6.52\varepsilon_0$). In (a): $a = 10\,nm$, $d_z = d_y = 22\,nm$, $d_x = 3d_y$; in (b) $a = 1\,nm$, $d_z = d_y = 2.2\,nm$.



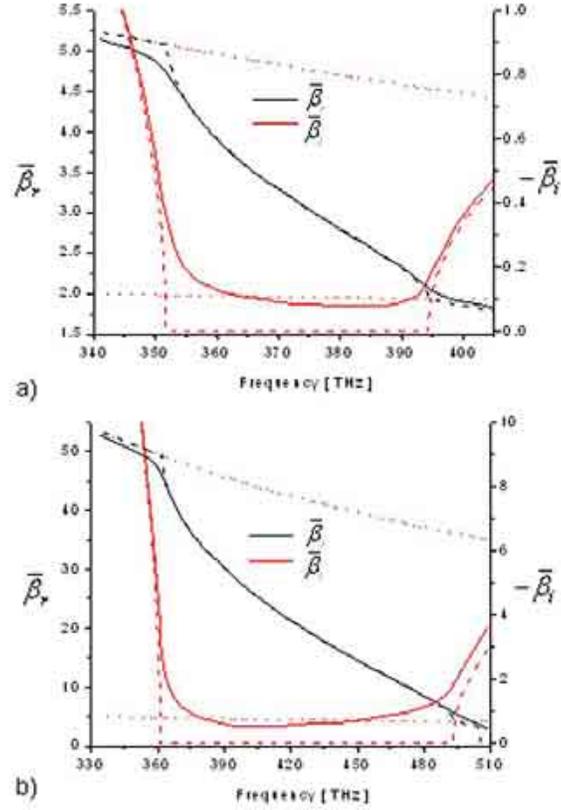

Figure 11 – (Color online). Similar to Fig. 9 and 10, but in a $Si$ background ( $\varepsilon_{Si} = 14.6\,\varepsilon_0$ ).In (a):

$a = 10\,nm$ , $d_z = d_y = 22\,nm$ , $d_x = 3d_y$ ; in (b) $a = 1\,nm$ , $d_z = d_y = 2.2\,nm$ .

In Figure 11, the same results are reported for a silicon background, which has

$\varepsilon_{Si} = 14.6\,\varepsilon_0$ . The results are consistent with the previous cases, but the resonance

frequency here has been moved down to the infrared regime, due to the different

contrast between silver and silicon. The resonances of the isolated particles is

located at $f_0 \simeq 375\,THz$ , $\lambda_0 \simeq 800\,nm$ . Employing $a = 10\,nm$ particles, the

normalized spacing in the backward region is $\overline{d}_y = 0.66$ , and this is shown in Fig.

11a, where again $d_z = d_y = 22\,nm$ and $d_x = 3d_y$ . Shrinking the size of the



particles to $a = 1\,nm$ and $d_y = 2.2\,nm$ again increases bandwidth and also the attenuation factors in this case. It is interesting to realize, however, that in this region the sensitivity to losses is lower, since the factor $\dfrac{\varepsilon_i \varepsilon_0}{\left(\varepsilon_r - \varepsilon_0\right)^2 + \varepsilon_i^2}$ in the expression (11) for $\overline{\alpha}_{loss}^{-1}$ is reduced, due to the more negative value of $\varepsilon_r$ in this range of frequency and the still relatively lower losses of silver in this frequency regime.

The previous numerical plots have considered a transverse spacing in the direction of the electric polarization three times larger than in the two other directions. This allows building a material that is isotropic only in the $y - z$ plane with a specific orientation of the electric field. However, if we want to design a fully 3-D isotropic metamaterial that is also capable of interacting with an arbitrary polarization of the field, we should consider a uniform spacing in all the three directions, as it is reported in the example of Fig. 12.

Fig. 12a shows the dispersion for an isotropic lattice of silver particles of radius $a = 5\,nm$ in a vacuum background with spacing $d_z = d_y = d_x = 11\,nm$. Here the range of frequency for the backward-wave propagation is shifted down with respect to the individual resonances of the particles, as predicted by the previous analysis. At the central frequency of this range, for this choice of parameters $\overline{d}_y = 0.18$, which is around the optimum point for relatively less sensitivity to losses in this configuration, as predicted by Fig. 8. However, in this isotropic case the realistic losses of silver are too high to support a propagating mode, and therefore just for the sake of exploring the dispersion of real and imaginary parts



of wave number, here we have considered the hypothetical value of $\text{Im}\,\varepsilon_{Ag} = 0.01$ in evaluating these results. The real part of permittivity still follows the realistic dispersion of silver. (We can consider other materials with lower losses, such as SiC and Bismuth, for these plasmonic particles forming the isotropic 3D metamaterials, but they would operate in the THz region.) Fig. 12b shows similar results, but using silicon as background, with spheres of radius $a = 4\,nm$ and $d_z = d_y = d_x = 8.8\,nm$. Clearly the negative-index propagation has been shifted down in frequency, yielding analogous results to Fig. 12a. In this case of isotropic lattice, the realistic imaginary part of the permittivity of silver from [48] is about 20 times larger than the one we considered in this last example, and therefore the damping factor would be proportionally higher for realistic silver loss, not allowing real propagation along the lattice. It should be underlined, however, that this absorption is mainly due to the high concentration of the field near and inside the particles, densely packed in the material. Employing a lower-loss material, or using this technique at different frequencies, may in principle allow the realization of isotropic backward-wave materials. The important point we can raise here is that the technique presented here is effectively adequate for designing a 3-D isotropic (and anisotropic) negative-index optical metamaterial. (If one wants to utilize silver for these plasmonic particles, the anisotropic 3D metamaterials discussed earlier in this section provide lower attenuation constant than in the isotropic case, and thus it would be more desirable. If, on the other hand, the isotropic case in this frequency range is the goal, one should not use silver for these particles. Instead, lower-loss materials such as SiC and Bismuth may be



more desirable.) A mathematical recipe to find the region in which the influence of material losses in this design is minimized has also been found. Better results are expected by employing plasmonic materials with lower values of the quantity

$\dfrac{\varepsilon_i \varepsilon_0}{\left(\varepsilon_r - \varepsilon_0\right)^2 + \varepsilon_i^2}$ at the frequency of interest.

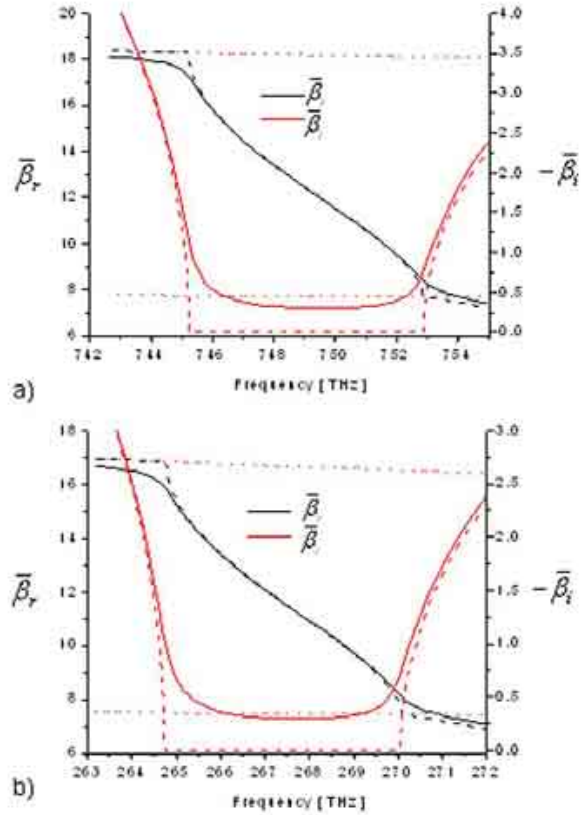

Figure 12 – (Color online). Similar to Fig. 9, 10 and 11. In (a): $a = 5\,nm$, $d_z = d_y = d_x = 11\,nm$ and the background is air; in (b) $a = 4\,nm$, $d_z = d_y = d_x = 8.8\,nm$ and the background is silicon. However, in these plots the imaginary part of permittivity of silver, used here, is not the realistic value, but instead a hypothetically lower imaginary part has been used to explore the variation of wave number in this case where losses affect more critically the wave propagation.



Before concluding this section, we underline that the analytical results reported in this manuscript have assumed that the permittivity of thin shells or tiny particles of plasmonic materials are those of the corresponding bulk materials. It is well known that such assumption is not necessarily adequate when extremely small plasmonic particles are considered [50]. Even though changes in the bulk material parameters are expected only when the thickness of the material samples is comparable with the mean free path of the conduction electrons in the material, i.e., a few nanometers, this may represent a limitation in the present analysis for the minimum size of the inclusions in such designs. Usually it should be considered that by scaling down the size of the particles to few nanometers the effective imaginary part of the plasmonic materials is increased. These effects are not of interest for the present manuscript, since the examples presented here are intended to show how these concepts are effectively applicable, and any deviations from the permittivity values that we have adopted here may not qualitatively modify the concepts behind the reported results.

## 10.    *Isotropy in 2-D and 3-D: Dispersion Diagrams*

The previous examples all relate to the case of TE propagation in one specific direction. The polarization of the electric field over all the particles was assumed to be directed along $x$, transverse to the direction of propagation $y$. It is obvious that in the cases where $d_z = d_y$, due to symmetry, the propagation along the $z$ direction would have the exact same properties, since the lattice is 2D isotropic. However, it is interesting to plot the whole dispersion plots in the transverse



plane, to study the propagation of TE plane waves in this medium even in oblique directions or when the medium is anisotropic. Again, due to the symmetry of the problem, the linear polarization along $x$ does not limit this analysis and similar results may be obtained for plane waves with other polarizations.

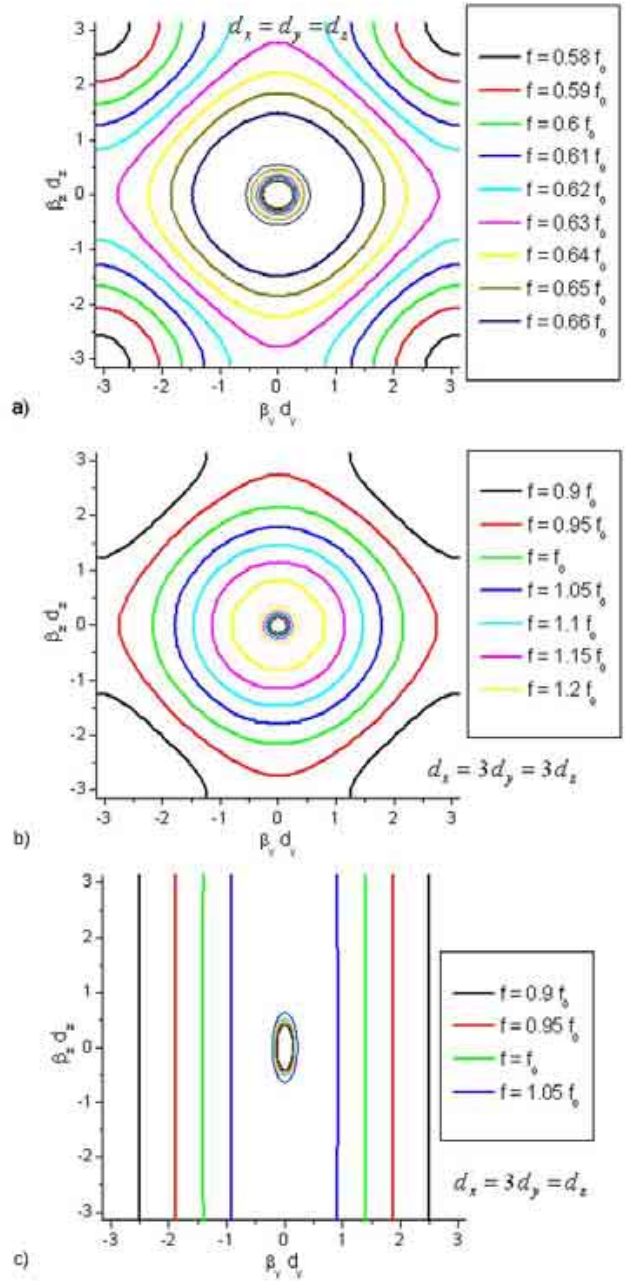



Figure 13 – (Color online). Equifrequency dispersion diagrams of the normalized wave vector $\overline{\boldsymbol{\beta}}$ in the transverse plane (with respect to the polarization of the field) for a metamaterial made of spherical particles with radius $a = \lambda_0 / 100$ (at the frequency $f_0$) and spacing $d_y = 2.1a$, with homogeneous permittivity for the particle following a Drude dispersion $\varepsilon(\omega) = \varepsilon_0 \left(1 - 3 f_0^2 / f^2\right)$.

Figure 13 shows the equifrequency dispersion curves for the TE modes ($\overline{\beta}_x = 0$) in the geometry of Fig. 6, i.e., for spheres with permittivity $\varepsilon(\omega) = \varepsilon_0 \left(1 - 3 f_0^2 / f^2\right)$, radius $a = \lambda_0 / 100$ at the frequency $f_0$ and center-to-center spacing equal to $d_y = 2.1a$. Fig. 13a shows the isotropic case, consistent with Fig. 6b. The results are clearly symmetric for $\overline{\beta}_y$ and $\overline{\beta}_z$. As seen from these figures, in the range of frequencies where $\overline{\beta}_{\min} < \overline{\beta}_y < \pi / \overline{d}_y$ with $\overline{\beta}_z = 0$, the propagation is backward and quasi-isotropic in the whole 2-D plane of transverse propagation. When the frequency reaches the point where $\overline{\beta}_y \simeq \pi / \overline{d}_y$ however, the dispersion curves are no longer exact circles, and for higher frequencies the propagation is allowed only in oblique directions, consistent with any periodic lattice (higher-order Floquet bands are of course available, but they are not considered here). It is noticeable how the backward-wave propagation coexists with forward-wave modes, arising at the center of the figure (for smaller values of $\overline{\beta}$), and maintaining the isotropic properties in this 2D transverse plane. In this case the material is isotropic in three directions, since changing the polarization of the electric field one obtains analogous plots in the other directions of propagation.



Increasing the spacing in the direction of polarization of the electric field, as in Fig. 13b, the bandwidth of the backward-wave operation increases notably, as already predicted in Fig. 6a. Still the propagation in the 2-D transverse plane is isotropic in a wide range of frequencies, and forward-wave modes are present around the center of the plots. As in the previous case, increasing the frequency the two families of curves (backward and forward) converge towards the circle for which $\bar{\beta}_y^2 + \bar{\beta}_z^2 = \bar{\beta}_{\min}^2$, at which the group velocity is zero in absence of losses. Everything is consistent with Fig. 6.

Increasing the spacing also in the transverse direction, i.e., along $z$, as shown in Fig. 6c produces, as expected, strong anisotropy in the transverse plane of propagation, with preferred $\beta_y$ directions, whose values are weakly dependent on the value of $\beta_z$. Again a family of forward-wave modes coexists with the backward-wave modes of propagation, sharing an analogous anisotropic behavior with eccentric ellipses. It is worth noting that in this last case the lattice is effectively composed of linear chains of densely packed particles that for sufficiently high $\bar{\beta}_y$, i.e., in their negative-index nano-TL operation, weakly interact with one another, since the modes are very confined around the particles composing the chains. In its backward mode of operation, therefore, such a medium behaves rather like a collection of parallel transmission lines, in many ways analogous to the wire medium in its canalization regime as proposed in [51] at microwave frequencies using conducting wires operating as transmission-lines or in its equivalent version made with plasmonic rods at infrared and optical frequencies [52]. Also the dispersion properties of Fig. 13c remind us of those



reported in [51]-[52], having a longitudinal wave number practically independent of the transverse component of the wave vector, even though here the supported modes are not necessarily TEM and therefore the wave number of propagation may be varied with frequency and does not depend only on the background material, but also on the properties of the particles composing the material. This effect confirms once again the nano-TL circuit analogy that represents the motivation behind this work. The supported modes in this configuration have backward-wave properties since, as we have shown in [34], such isolated chains would transport highly confined backward waves in their transversely-polarized propagation. Several interesting imaging applications, proposed for the wire medium at microwaves, may be extended to optical frequencies with the present setup, analogous with what was proposed in [52] in studying propagation along parallel plasmonic nano-rods. In our case, the role of the nano-wires is taken by the plasmonic nano-TL chains supporting tightly confined modes. Here the chains may provide even further degrees of freedom. We are currently studying some of these exciting possibilities.

As a preliminary result, in Fig. 14 we report a full-wave simulation obtained with finite-integration-technique full-wave commercial software [53] for the case of a metamaterial slab composed of $7 \times 7 \times 21$ silver nanoparticles of radius $a = 5\,nm$ at the frequency $f_0 = 600\,THz$ ($\lambda_0 = 500\,nm$), with lattice constants $d_x = d_y = 2\,d_z = 22\,nm$ (notice that in this simulation the propagation direction is along the $z$ axis). In the simulation realistic values of permittivity for sliver has been employed at the frequency of interest. At the entrance face of the slab, in the



$x - y$ plane, two short electric dipoles, separated by $44 \, nm$ from each other, excite the nano-TL mode that propagates along the set of parallel chains. Similar to the wire medium utilized for similar purposes, the guidance properties of such chains allow "transferring" an image from the entrance to the exit face with its sub-wavelength details preserved. This is clear from the electric field distribution in the $xz$ plane (Fig. 14a), which shows how the electric field distribution (brighter colors correspond to higher amplitudes of the field) travels from the entrance to the exit face of the slab. In Fig. 14b the two point-like sources are resolved at the exit face of the slab, even though their distance is just less than a tenth of free-space wavelength. Notice that this sub-wavelength resolution is possible thanks to the flat response of the dispersion diagram of Fig. 13c in the transverse plane, implying that the spectrum of plane waves composing the image is uniformly "canalized" by the nano-TL imager from the entrance to the exit face. As in the other canalization devices proposed in the literature [51]-[52], the fundamental bottom limit to the resolution of this device is represented by the transverse distance among the nanochains, which in this case is of $44 \, nm$.

As an aside, it is interesting to point out how these analogies between the metamaterial presented here and the nano-rod medium in [52] (which in many ways is an extension of the wire medium [54] to optical frequencies) are confirmed by the strong spatial dispersion arising in the metamaterial under analysis here, which is responsible for the propagation of two TE modes in the same frequency band, as noticed in Section 7. This is consistent with the previous results and appears to be an intrinsic property of such materials with densely



packed long inclusions. The nano-TL modal operation, in fact, may be viewed as the origin of the strong non-locality of the lattice in its entirety, similar to what happens in the wire media or analogous metamaterials where electrically long conducting inclusions support TL modes that can electrically connect distant points in space. Here their role is taken by the nanochains of plasmonic particles composing the lattice, which in [34]-[35] have been shown to support similar TL propagation.

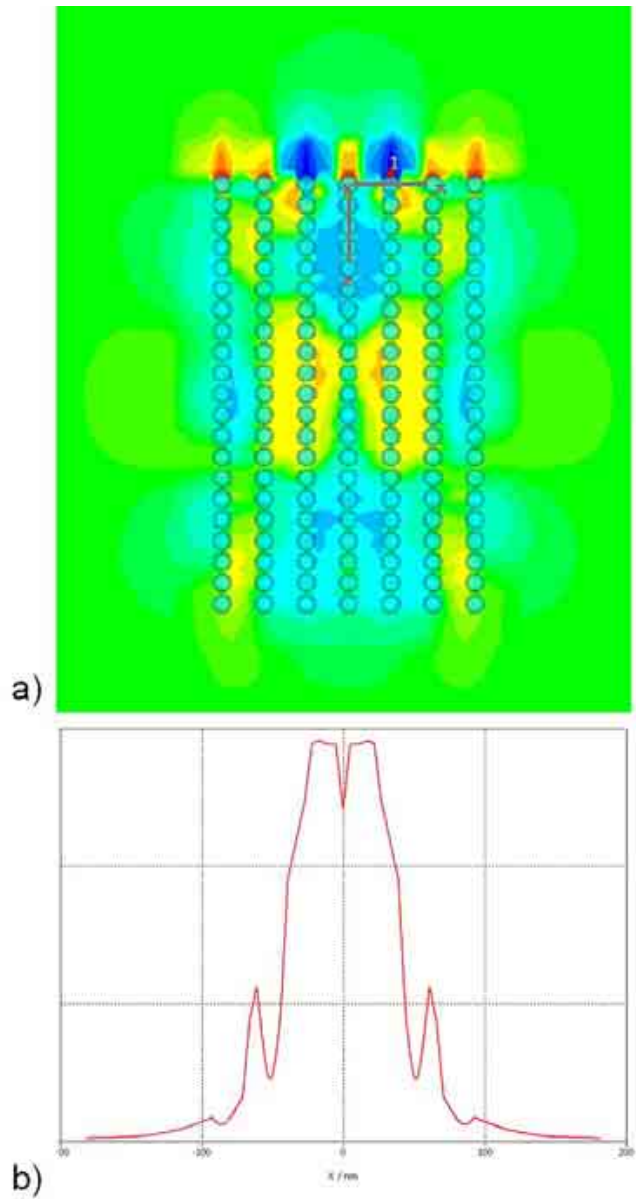



Figure 14 – (Color online). Sub-wavelength imaging in the canalization regime for a 3D metamaterial slab made of silver particles with $d_x = d_y = 2 d_z = 22\,nm$. The total dimensions of the slab are $7 \times 7 \times 21$ nanoparticles and two short electric dipole sources, separated by $44\,nm$ are placed in front of the entrance face. In (a) the electric field amplitude on the $x - z$ plane is reported, where brighter colors correspond to higher intensity of the field. In (b) the electric field distribution (arbitrary units) at the exit face of the slab is also reported, showing the sub-wavelength resolution due to the canalization phenomenon.

The case of $3 d_x = 3 d_y = d_z$, which would correspond to the case of Fig. 13b when the polarization of the field is rotated, would support modes around the same band of frequencies of the isotropic case (Fig. 13a), as seen from the dispersion plots of Fig. 2. Their dispersion in the 2-D transverse plane, not reported here, would be similarly anisotropic as the one of Fig. 13c. This implies that such a lattice with non-uniform lattice periods in the transverse plane would respond differently to the two orthogonal polarizations of the electric field, with distinct pass-bands and transmission response. Also this may have some interesting applications, since such a lattice, basically composed of parallel planes of densely packed particles may act as a 2-D DNG metamaterial for one polarization and as a canalization imager for the other polarization, with the longitudinal component of the guided $\beta$ being weakly dependent on its transverse component. Notice how all the confined modes described here are backward in nature, and we should consider that the strong spatial dispersion of the lattice implies the simultaneous presence of forward modes for smaller $\beta$. We argue that such modes may be eliminated with some conventional techniques, i.e.,



adding resistive sheets at the middle planes among the particles, where the tightly confined backward modes are less present, or with a proper polarization of the exciting field. This is currently under investigation. Also the dual configuration of dielectric lattices in a plasmonic background would allow to suppress this second mode of propagation, since in this case the background material by itself does not support any propagating modes. In a recent symposium, we have presented orally some preliminary results in this sense in [55] and we are currently working on these problems to further explore these properties.

It is worth noting that in this manuscript we have concentrated our attention on the propagation transverse to the direction of polarization of the electric field, which is the one consistent with our TL modal analysis. In principle, however, as previously noticed, propagation with wave vector having a non-zero component along the direction of the electric field is also possible, as it happens at the plasma frequency in plasmonic materials or in a wider range of frequencies in spatially dispersive crystals [46]. This is also predicted by the NNA in Eq. (3) and it would be consistent with the longitudinal propagation in the 1-D chain problem [34] and with the other analogous setups we have studied [19], [41]. We will analyze in more details this set of modes in a future work, but here for completeness we report in Fig. 15 the dispersion in the $\beta_x - \beta_y$ plane, which, when combined with the previous diagrams, provides a full understanding of the behavior of the propagating modes in this medium. This is done for the two representative cases of Fig. 13a and Fig. 13c.



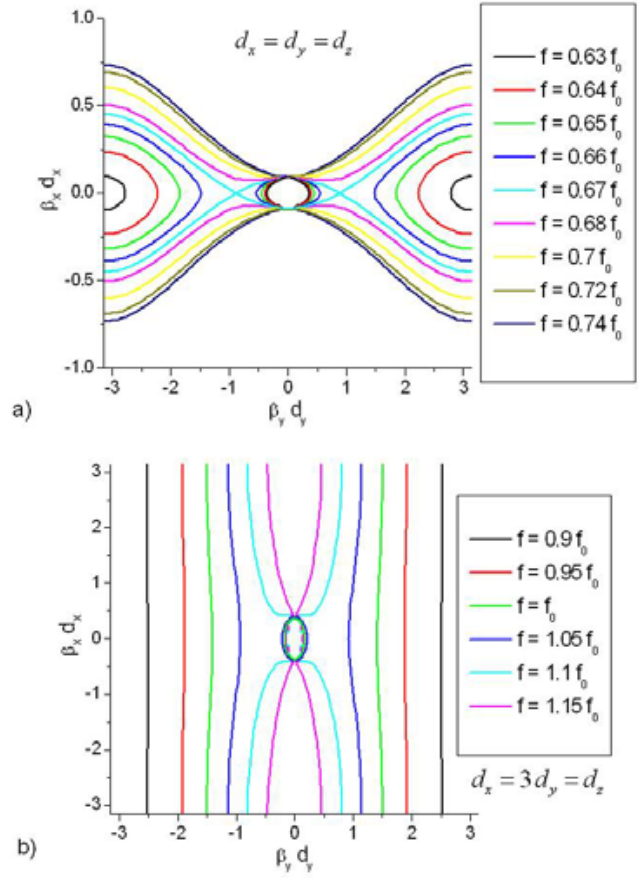

Figure 15 – (Color online). Equifrequency dispersion curves of the normalized wave vector $\overline{\boldsymbol{\beta}}$ in the longitudinal plane (longitudinal with respect to the polarization of the field) for a metamaterial made of spheres with radius $a = \lambda_0/100$ and spacing $d_y = 2.1a$ at the frequency $f_0$, with homogeneous permittivity of the particles following a Drude dispersion $\varepsilon(\omega) = \varepsilon_0\left(1 - 3f_0^2/f^2\right)$.

Fig. 15a refers to the isotropic case ($d_x = d_y = d_z$). As can be seen from this figure, in the negative-index region the plots have a hyperbolic shape in this plane, and the phase vector may have a longitudinal component, even though limited to relatively small values. All the curves converge to the point where $\left(\overline{\beta}_x \simeq 1,\ \overline{\beta}_y \simeq 0\right)$, for which the mode is longitudinal and analogous to a



transcription-line regime, with the wave number being very close to the one of the background material. At the frequencies for which the backward-wave propagation is possible, also forward modes are present with elliptical dispersion curves in this plane. At the frequency $f = 0.67 f_0$ the two modes merge on the $\bar{\beta}_y$ axis at the point $\bar{\beta}_y = \bar{\beta}_{min}$, as already predicted by the previous discussion.

When the medium becomes a set of parallel chains, i.e., for larger spacing among the chains, as in Fig. 15b ($d_x = 3d_y = d_z$), on the other hand, the canalization regime is evident also in this dispersion plane, with the supported $\bar{\beta}_y$ weakly dependent on $\bar{\beta}_x$. Forward modes are still present for lower $\bar{\beta}$ also in this plane.

## 11.    Conclusions

In this paper, we have applied the concepts of nanocircuit elements at optical frequencies to design a 3-D plasmonic optical nano-TL metamaterials with relatively broad bandwidth and isotropic properties, and we have provided a detailed theory for such 3-D optical nano-TL metamaterials. This class of materials may act as a negative-index material at infrared and optical frequencies. The optimum conditions on bandwidth and losses have been derived analytically and full-wave analytical results have been provided to characterize the material in its negative-index operation in the 3-D space. Strong spatial dispersion, analogous to other transmission-line materials at lower frequencies are predicted, but this does not affect the main conclusions on the realistic possibility of obtaining a plasmonic metamaterial with backward-wave properties. Applications for sub-



wavelength imaging at optical frequencies have been envisioned with this setup at optical frequencies.

**Acknowledgements**

This work is supported in part by the U.S. Air Force Office of Scientific Research (AFOSR) grant number FA9550-05-1-0442. Andrea Alù was partially supported by the 2004 SUMMA Graduate Fellowship in Advanced Electromagnetics.

Italy, March 28-31, 2004. One page abstract is in the CD digest of abstracts.